\documentclass[11pt]{article}
\usepackage{amsmath}
\usepackage{graphicx,psfrag,epsf}
\usepackage[utf8]{inputenc}
\usepackage{enumerate}
\usepackage[dcucite]{harvard}  
\usepackage{caption}
\usepackage{subcaption}
\usepackage{float}
\usepackage{multicol}
\usepackage{booktabs,siunitx}
\usepackage{pdfpages}
\usepackage{multirow}
\usepackage{adjustbox}
\usepackage{algorithm}
\usepackage{algorithmic}
\usepackage{url}

\newcommand{\blind}{0}

\newcommand{\be}{\begin{equation}} 
\newcommand{\ee}{\end{equation}}

\newcommand{\bq}{\begin{equation}} 
\newcommand{\eq}{\end{equation}}

\newcommand{\bc}{\begin{center}} 
\newcommand{\ec}{\end{center}} 
\newcommand{\bi}{\begin{itemize}} 
\newcommand{\ei}{\end{itemize}} 
\newcommand{\VEKK}[1]{}

\newcommand{\vekk}[1]{}

\newcommand{\RNum}[1]{\uppercase\expandafter{\romannumeral #1\relax}}
\newcommand{\squarebrackets}[1]{\left[#1 \right]}

\let\citeN=\citeasnoun

\begin{document}

\def\spacingset#1{\renewcommand{\baselinestretch}%
{#1}\small\normalsize} \spacingset{1}


\if1\blind
{
  \title{ Monitoring time to event in registry data using CUSUMs based on relative survival models}
  \author{Jimmy Huy Tran\footnote[1]{Department of Mathematics and Physics, 
          University of Stavanger, Norway. E-mail: jimmy.tran@uis.no. Corresponding author.},  
          Jan Terje Kval\o y\footnote[2]{Department of Mathematics and Physics, 
          University of Stavanger, Norway. E-mail: jan.t.kvaloy@uis.no.}
          and Hartwig Kørner\footnote[3]{
          Department of Gastrointestinal Surgery, Stavanger University Hospital Stavanger, Norway and 
Department of Clinical Medicine, University of Bergen, Bergen, Norway. E-mail: Hartwig.Korner@uib.no}}
\date{}
} \fi

\if0\blind
{
  \begin{center}
    {\Large\bf Monitoring time to event in registry data using CUSUMs based on relative survival models}

  \bigskip \bigskip
  Jimmy Huy Tran\footnote[1]{Department of Mathematics and Physics, 
          University of Stavanger, Norway.  Corresponding author. E-mail: jimmy.tran@uis.no.},  
          Jan Terje Kval\o y\footnote[2]{Department of Mathematics and Physics, 
          University of Stavanger, Norway.}
          and Hartwig Kørner\footnote[3]{
          Department of Gastrointestinal Surgery, Stavanger University Hospital Stavanger, Norway and 
Department of Clinical Medicine, University of Bergen, Bergen, Norway.}\\[2mm]
\medskip
\today
\end{center}
} \fi

\medskip
\bigskip
\begin{abstract}

An aspect of interest in surveillance of diseases is whether the survival time distribution changes over time. By following data in health registries over time, this can be monitored, either in real time or retrospectively. With relevant risk factors registered, these can be taken into account in the monitoring as well. A challenge in monitoring survival times based on registry data is that the information related to cause of death might either be missing or uncertain. To quantify the burden of disease in such cases, relative survival methods can be used, where the total hazard is modelled as the population hazard plus the excess hazard due to the disease. 

We propose a CUSUM procedure for monitoring for changes in the survival time distribution in cases where use of excess hazard models is relevant. The CUSUM chart is based on a survival log-likelihood ratio and extends previously suggested methods for monitoring of time to event data to the excess hazard setting. The procedure takes into account changes in the population risk over time, as well as changes in the excess hazard which is explained by observed covariates. Properties, challenges and an application to cancer registry data will be presented.

\end{abstract}

\noindent%

   
\section{Introduction}
\label{sec:intro}

In health registries, such as cancer registries, patients are routinely registered at diagnosis of disease, and outcome data like survival times are often added later. Based on such data, an aspect of interest could be to monitor whether the distribution of the time to an outcome of interest changes over time, for instance if the survival times of cancer patients due to the disease itself change over time while adjusting for known risk factors. Such monitoring could be of interest both for real-time monitoring of incoming data and for retrospective analyses to pinpoint when in time important changes took place. 

A common challenge in monitoring survival times based on such registry data is that time to death, but not necessarily cause of death, is registered. In addition, even if this information is available, it is often unreliable at population level. Therefore, to quantify the burden of disease in such cases, relative survival methods can be used. Under this framework, the frequent standard assumption is that the total hazard is modelled as the population hazard plus the excess hazard due to the disease, see e.g. \citeN{dickman2004regression}, \citeN{perme2009approach} and \citeN{perme2012estimation} for an overview of notions in relative survival and regression models regarding this topic.  The population hazard is usually retrieved from national population life tables, stratified on demographic variables.

Methods from statistical process control have been essential in many different fields and applications, originally suggested for monitoring processes in various industries. One of the most widely used control charts is the CUSUM (cumulative sum) presented by \citeN{page1954continuous}. Compared to other charts, e.g. Shewhart charts, the CUSUM chart is known to perform better for detecting smaller shifts that are persistent over time, thus being suitable for a range of medical settings. Over the years, extensions and adaptions of CUSUM charts to a number of different scenarios have been developed. CUSUM procedures for monitoring of ordinary time to event data were first proposed by \citeN{Biswas2008Riskadjusted}, \citeN{Sego2009Riskadjusted}, \citeN{Gandy2010ram} and \citeN{Steiner2010EWMAsurv}. These have since been extended to 
contexts like frailty models \cite{Begun2019CUSUMfrailty,Keshavarz2021CUSUMfrailty},  
cure models \cite{Oliveira2016Longterm},
illness-death models \cite{Liu2023SCRCUSUM} and queue models \cite{Kuang2023CUSUMqueues}. 
Impact of estimation error in time to event data monitoring was studied by \citeN{Zhang2016Estimationerror} and effectiveness versus periodic evaluations was studied by \citeN{Massarweh2021}. Applications have been demonstrated in, for instance, monitoring of perioperative mortality \cite{Lai2021CUSUMsurgerysurvival,Chen2023CUSUMperiopmortality}.

We propose a CUSUM procedure for monitoring for changes in the time to event distribution when the use of excess hazard models is relevant, e.g. for monitoring based on registry data with uncertain or missing cause of death information, a typical example being cancer registry data. The CUSUM chart is based on a survival log-likelihood ratio and extends the literature discussed above on monitoring of time to event to the relative survival setting. The procedure takes into account changes in the population risk over time, as well as changes in the excess hazard which are explained by observed covariates. Properties, challenges and an application to cancer registry data will be presented.

The structure of the paper is as follows: Section \ref{sec:CUSUM} introduces the set-up and notation and presents the proposed method. In Section \ref{sec:sim}, numerical simulations and experiments are carried out to demonstrate the use of the method and how it performs under different scenarios, including studying the impact of estimation error. An application of the method on a real data set obtained from the Norwegian Cancer Registry is illustrated in Section \ref{sec:realdata}. Finally, Section \ref{sec:conc} provides some concluding remarks. R-code implementing the proposed methods and for running the simulations in Section \ref{sec:sim} can be found at \url{https://github.com/jihut/cusum_relative_survival_simulations}. 

\section{CUSUM chart based on excess hazard models in relative survival setting}
\label{sec:CUSUM}

\subsection{Set-up and Notation}
\label{subsec:notation}

The setting considered is monitoring time to event under the assumption of an additive hazard model  $h(\cdot)=h_P(\cdot)+h_{E}(\cdot)$. Here, $h_E(\cdot)$ denotes the excess hazard due to the disease of interest. In theory, $h_P(\cdot)$ represents the hazard related to other causes. However, it is almost always the case that this quantity is substituted by the population hazard, which is assumed to be known and usually extracted from population life tables. 
We would like to monitor for changes in $h_E(\cdot)$ over calendar time. 
 
For ease of notation, consider a monitoring system starting from a given calendar date which is now defined as the origin time point $t=0$ of the monitoring period, for instance the first day of a particular year. Let  $0 \leq B_1\leq B_2 \leq \dots $ denote the arrival times after the starting point at which individuals enter the monitoring system. This could for instance be when a patient is diagnosed and enters the cancer registry system. 
Let $Z_1, Z_2, \ldots$ denote the corresponding vectors of demographic variables determining $h_P(\cdot)$. Age, gender and birth year are always included as a part of $Z_i$, but other variables could exist as well depending on different population life tables. Similarly, $X_1, X_2, \dots $ represent the covariate vectors affecting $h_E(\cdot)$. Usually, $X_i$ contains available and relevant information regarding a patient, e.g. cancer stage and treatment, where $Z_i$ may be a subset of $X_i$. 

 Further, denote $T_{Ei}$ as the time to event due to the disease of interest. Here, the event of interest will be death. Similarly, $T_{Pi}$ represents the time to event due to other causes appearing in the general population. With the additive hazard model and the assumption of $T_{Ei}$ and $T_{Pi}$ being conditionally independent given the covariates (\citeN{perme2012estimation}), let $T_i = \text{min}(T_{Ei}, T_{Pi})$ denote the overall time to event and $C_i$  the censoring time for individual $i$. 
 We do not observe the cause of event, only $\min(T_i, C_i)$ and whether the observation was censored or not. We then define the time at risk up to time $t$ after the start calendar time of the monitoring system for individual $i$ as  $A_i(t) =\min(T_i,C_i,\max(t-B_i,0))$ and the event indicator $\delta_i(t)=I(T_i= A_i(t))$.  

Let $h_{0i}(\cdot)=h_P(\cdot,Z_i)+h_{E,0}(\cdot,X_i)$ be the \emph{in-control} hazard rate, i.e. the reference (or baseline) hazard based on the current situation at the start of the monitoring. This is the situation we want to be able to quickly detect deviations from.  The known population hazard $h_P(\cdot)$ might change over calendar time, but this is suppressed in the notation. The excess hazard rate in the in-control (baseline) situation is assumed to be constant over calendar time. At some time   $\eta$ after the monitoring start, the hazard switches to some \emph{out-of-control} hazard rate $h_{1i}(\cdot)=h_P(\cdot,Z_i)+h_{E,1}(\cdot,X_i)$, i.e. a hazard rate that deviates from the reference period, and we would like to quickly detect this change. Finally, let the cumulative hazard be $H_{ji}(t)=\int_0^t h_{ji}(s)ds$ for $j=0,1$.

\subsection{The proposed CUSUM procedure}
\label{subsec:CUSUM} 

Following \citeN{Gandy2010ram}, we define a new continuous time CUSUM procedure for the setting of relative survival based on the survival likelihood ratio. \citeN{Moustakides1986} has shown optimality properties for CUSUMs based on likelihood ratios.  The (partial) likelihood based on $h_{ji}$  up to time $t$ after the start of monitoring is \cite{abg2008}
$$
L_j(t)=\prod_{i: B_i\leq t} h_{ji}(T_i)^{\delta_i(t)}\exp[-H_{ji}(A_i(t))].
$$
The (partial) log-likelihood ratio between the in- and out-of-control likelihood up to time $t$ thus becomes
$$
R(t)=\sum_{i: B_i\leq t} \delta_i(t)\log\left(\frac{h_{1i}(T_i)}{h_{0i}(T_i)}\right)
-\sum_{i: B_i\leq t}  \left[H_{1i}(A_i(t))-H_{0i}(A_i(t))\right].
$$
$R(t)$ is the continuous time analogue to a cumulative sum. By restarting this sum whenever it drops to 0, we get the continuous time CUSUM control chart 
$$
\Psi(t) = R(t) - \min_{s\leq t} R(s)
$$
that signals at a time $\tau=  \inf\{t: \Psi(t)>c\}$ for some threshold $c>0$. 

What differs from the procedure in \citeN{Gandy2010ram} is that we here work with an excess hazard model and are interested in the change of $h_E(\cdot)$, and this in particular implies that it is more challenging to determine $c$ and that we need to keep track of the population hazard in the first term of $R(t)$. Usually, the population hazard can be found in national life tables. Also, in most practical applications, one has to estimate $h_{E,0}(\cdot,X_i)$. A popular excess hazard model in relative survival is the proportional excess hazard model, i.e. $h_{E,0}(\cdot,X_i) = h_0(\cdot) \exp(\mathbf \beta X_i)$, where $h_0$ corresponds to the baseline excess hazard. For instance, all the available models implemented in the popular R package for relative survival \texttt{relsurv} \cite{perme2018nonparametric} inherit this assumption, as well as models in Stata \cite{dickman2015Stata}.
For our purposes, a piecewise constant baseline excess hazard model, estimated by a GLM-approach with Poisson error structure \cite{dickman2004regression}, and a semiparametric excess hazard model with a smooth nonparametric estimate of $h_0(\cdot)$, fitted by an EM-algorithm \cite{perme2009approach}, will be the relevant models in the upcoming sections.

\subsection{Out-of-control alternatives}
\label{subsec:outofcontrol} 
In this section, we consider different alternative models for the out-of-control hazard $h_{1i}(\cdot)$. As motivated in the previous sections, the main focus here is to monitor the change in the excess hazard part related to the disease of interest. Therefore, the alternative considered in \citeN{Gandy2010ram}, which implies monitoring change in the total hazard, is not suitable for the current purpose. However, based on the same idea, a proportional alternative directly on the excess hazard, i.e. $h_{E,1}(\cdot,X_i)=\rho h_{E,0}(\cdot,X_i)$ for some $\rho > 0$, can be considered. The corresponding log-likelihood ratio after a monitoring time period of length $t$ becomes 
\begin{equation}
R(t)=\sum_{i: B_i\leq t} \delta_i(t)\log
\left(\frac{h_P(T_i,Z_i)+\rho h_{E,0}(T_i,X_i)}{h_P(T_i,Z_i)+h_{E,0}(T_i,X_i)}\right)
-(\rho-1)\sum_{i: B_i\leq t}  
\left[ H_{E,0i}(A_i(t))\right].
\label{eq:loglik_prop}
\end{equation}
Here, $\rho$ is the proportional factor between the out-of-control and in-control hazard. If $\rho > 1$, one can interpret this as a situation where the out-of-control hazard has increased by $(\rho - 1) \cdot 100\%$ compared to the reference hazard. Similarly, if $\rho < 1$, the burden of disease has been decreased by $(1 - \rho) \cdot 100\%$ with respect to the in-control hazard. 

The proportional alternative implies assuming a larger change in absolute value of the hazard for patients with a higher hazard. However, this might not always be reasonable, and an alternative model for the change could be an additive model where the hazard changes by the same absolute amount for all patients.  This motivates an additive out-of-control alternative for the excess hazard. Assuming that we work with non-negative excess hazard, which is usually the case when dealing with cancer patients, the additive alternative is given as $h_{E,1}(\cdot,X_i) = \text{max}(0, h_{E,0}(\cdot,X_i)+\gamma) $ for some $-\infty < \gamma < \infty$ so that $H_{E,1}(t,X_i) = \int_0^t \text{max}(0, h_{E,0}(\cdot,X_i) + \gamma) du$ and 
\begin{align}
    R(t)&=\sum_{i: B_i\leq t} \delta_i(t)\log
\left(\frac{h_P(T_i,Z_i)+\text{max}(0, h_{E,0}(T_i,X_i)+\gamma)}{h_P(T_i,Z_i)+h_{E,0}(T_i,X_i)}\right) 
\nonumber
\\
&-\sum_{i: B_i\leq t} \left(\left( \int_0^{A_i(t)} \text{max}(0, h_{E,0}(u,X_i) + \gamma) du \right) - H_{E,0}(A_i(t),X_i) \right).
\label{eq:loglik_add}
\end{align}
The additive change parameter $\gamma$ can therefore be interpreted as the change in the excess hazard in absolute measure. If $\gamma <0$, the out-of-control hazard is smaller than the in-control, which implies that the disease mortality has been reduced. On the other hand, $\gamma >0$ represents an increase of the excess hazard. Here, the use of the maximum function is to ensure that we do not get a negative excess hazard in the out-of-control scenario for some individuals when $\gamma < 0$. In the case where $\gamma > 0$, the maximum function can be omitted and the log-likelihood ratio simplifies to 
\begin{align}
 R(t)&=\sum_{i: B_i\leq t} \delta_i(t)\log
\left(\frac{h_P(T_i,Z_i)+h_{E,0}(T_i,X_i)+\gamma}{h_P(T_i,Z_i)+h_{E,0}(T_i,X_i)}\right)
-\gamma\sum_{i: B_i\leq t}  
 A_i(t).   
\label{eq:loglik_add_without_max}
\end{align}
The set-up can in theory be extended to negative excess hazard as well by rather requiring that the total hazard is non-negative, i.e. $h_{1i}(\cdot) = \text{max}(h_P(\cdot,Z_i)+ h_{E,0}(\cdot,X_i)+\gamma, 0)$. For situations where $h_{1i}(T_i) = 0$, the corresponding observations will have no events. The hazard term can thus be omitted for these. However, we will not pursue this somewhat odd situation any further. 

Further, similar as considered in \citeN{Gandy2010ram} for ordinary survival models,  time-transformation alternatives can be used to model changes in the hazard. This is for instance done in order to incorporate more general changes that depend on the definition of the hazard function, e.g. non-proportional change between the in-control and out-of-control hazard. We consider the following linear accelerated time alternative specified as $H_{E,1}(u,X_i) = H_{E,0}(k u,X_i)$ for some $k > 0$. This leads to the following log-likelihood ratio for the linear accelerated time alternative:
\begin{equation}
R(t)=\sum_{i: B_i\leq t} \delta_i(t)\log
\left(\frac{h_P(T_i,Z_i)+k h_{E,0}(k T_i,X_i)}{h_P(T_i,Z_i)+h_{E,0}(T_i,X_i)}\right)
-\sum_{i: B_i\leq t}  
\left[H_{E,0i}(k A_i(t)) - H_{E,0i}(A_i(t))\right].
\label{eq:acc_lin_time}
\end{equation}
This parameterization implies in most cases an easy interpretation of the alternative - a value of $k < 1$ yields $H_{E,1i}(u) = H_{E,0i}(k u) < H_{E,0i}(u)$, which implies that the survivor function is larger at the same time point $u$ in the out-of-control setting compared to the in-control scenario, and the other way around for $k>1$. In other words, $k > 1$ accelerates time, indicating a faster increase in the cumulative excess hazard so that the out-of-control excess survival is smaller than the in-control period. In contrast, $k < 1$ slows time down and can be interpreted as an improvement in the burden of disease. One could also choose the parameterization $h_{E,1}(u,X_i) = h_{E,0}(k u,X_i)$, but the interpretation is less clear as it will in general depend on the shape of $h_{E, 0}(u, X_i)$ as a function of $u$ and thus will not be considered further here. 

A possible challenge with the linear accelerated time alternative occurs if the monitoring system is used to monitor survival up to a certain time point, for instance 10 year survival, and a $k>1$ alternative is of interest. Then, estimation of the hazard function up to  $10k$ years will be required for calculating $h_{E,0}(k u,X_i)$ and $H_{E,0}(k u,X_i)$. If nonparametric modelling is used, this requires that survival data up to $10 k$ years must be available for estimating the hazard function. If survival data beyond 10 years are not available, a parametric hazard model can be used, but this will require extrapolations beyond the range of the data used to estimate the model. 

A final note is that in practical settings, one needs to choose a value of $\rho$, $\gamma$ or $k$ in order to calculate the charts. Unless a clear specification of a single alternative of interest is given, an option is to set up several charts to monitor varying degrees of change either in the negative or positive direction of excess mortality. This is done as an illustration in the cancer data example in Section \ref{sec:realdata}. 

\subsection{Criteria for determining the signal threshold}
\label{subsec:threshold} 
The signal threshold $c$ needs to be tuned to obtain the desired performance of the CUSUM charts. Usually, the threshold is chosen so that the charts achieve a certain performance under the in-control alternative. For example, one could relate $c$ with the in-control average run length (ARL) defined as $\text{ARL} = E(\tau \mid \eta = \infty)$, which corresponds to the expected time until the charts cross the threshold $c$ when the hazard never changes to the out-of-control state. Then, $c$ is chosen such that the in-control ARL is equal to a desired value. Another common strategy is to choose $c$ such that the probability of a false signal until a given time point $t_m$ is at a desired level $\alpha$, i.e. $p_{\text{hit}} = P(\tau \leq t_m \mid \eta = \infty) = \alpha$. This is often called the in-control hitting probability. 

\citeN{Gandy2010ram} explored the setting of choosing $c$ with respect to the expected number of events until hitting or the probability that the number of events at hitting time point does not exceed a given amount when in-control, i.e. $E(N(\tau) \mid \eta = \infty)$ and $P(N(\tau) \leq N_\text{max} \mid \eta = \infty)$. In the ordinary survival model setting studied in \citeN{Gandy2010ram}, if a proportional change scenario is considered, $c$ can be analytically computed using a method based on a discrete time Markov chain. However, this methodology does not apply for the setting of monitoring the excess hazard as the jump of the chart is no longer constant across observations. Nevertheless, for the proportional change alternative, it is possible to approximate the current setting to a situation where the Markov chain method is applicable if the population hazard is much smaller than the excess hazard. This was examined in the master thesis of \citeN{Tran2022}, where the thresholds obtained from the method yield approximately the desired performances in circumstances where the excess hazard dominates the population hazard. 
Otherwise, for general excess hazard scenarios studied in the present paper, it is seemingly not possible to compute $c$ analytically. In the remaining sections, the thresholds will thus be calculated via different simulation approaches depending on the setting. 

\subsection{Simulating the signal threshold}
\label{subsec:simthreshold}

Motivated by the preceding section, we opt for the standard strategy of simulating the threshold so that a desired value of the false signal probability (in-control hitting probability) is achieved. In the following, an overview of the general approach to simulate the threshold $c$ based on a predefined $p_\text{hit}$ is described. 
\begin{algorithm}
    \caption{Simulating $c$ based on a chosen $p_\text{hit}$}
    \begin{algorithmic}[1]
        \STATE Specify the start time of the monitoring process, $t_m$, $p_\text{hit} = \alpha$ for some $\alpha \in (0, 1)$ and the number of simulations $N$. Assume that the arrivals of observations follow a homogeneous Poisson process with intensity $\lambda_a$ and the interim censoring time $C_i^*$ an exponential distribution with censoring rate $\lambda_{C^*}$.
        \FOR{$j = 1, ..., N$}
        \STATE Simulate the number of observations $n$ up to the time point $t_m$ with the assumed arrival process. Obtain $B_1, \cdots, B_n$. 
        \FOR{$i = 1, ..., n$}
        \STATE Simulate $X_i$ and $Z_i$. 
        \STATE Simulate $T_{Ei}$ based on $h_{E,0}(\cdot,X_i)$ and $T_{Pi}$ based on $h_{P}(\cdot,Z_i)$.
        \STATE Simulate interim censoring time $C_i^* \sim \text{Exp}(\lambda_{C^*})$. Calculate final censoring time $C_i = \text{min}(C_i^*, t_m - B_i)$. 
        \STATE Calculate $T_i = \text{min}(T_{Ei}, T_{Pi}, C_i)$.
        \ENDFOR
                \STATE Run CUSUM chart and let $W_j = \text{max}_{0 \leq t \leq t_m} \Psi_j(t)$, where $\Psi_j(t)$ is the CUSUM chart run for this specific iteration. 
        \ENDFOR
        \STATE Calculate $c$ as the $\alpha\cdot100$\% upper quantile of the values $W_1, \cdots, W_N$. 
    \end{algorithmic}
\end{algorithm}

We will consider two different scenarios for simulating the covariates $X_i$ and $Z_i$. For the first simulations performed in Section \ref{sec:sim}, we assume that the true covariate distribution is known and therefore simulate $X_i$ and $Z_i$ directly from these distributions. In real-life applications, this is usually not the case. We then opt for a nonparametric bootstrap procedure that estimates the baseline/in-control distribution in such scenarios. Also, the interim censoring rate is estimated from baseline data.

\subsection{Monitoring and updating schemes}
\label{subsubsec:updating}

The proposed method is best suited for detecting changes in the setting where all observations in the system will experience the out-of-control hazard whenever $t > \eta$, e.g. if a new regime for post-operative treatment is introduced for all patients.  However, in certain cases, only the individuals arriving after the change point $\eta$ will be affected by the change in the hazard, for instance if a new surgery procedure is introduced. Our method should still be able to detect the change in this latter scenario, but there will be more delay in detecting the change.

 With the definition $A_i(t) =\min(T_i,C_i,\max(t-B_i,0))$, we assume that information on the status of a patient is available continuously, from arrival and onwards. This is realistic if cases are registered to the database more or less in real time and information about censoring or events is swiftly added. For instance, this will be the case when the method is used for retrospective analysis.  Another scenario could be that information about each case becomes available only after censoring or an event. For this setting, \citeN{Gandy2010ram} suggested to redefine $A_i(t)$ such that the observation contributes to the likelihood ratio only after an event or censoring has occurred: 
\begin{equation*}
    A_i(t) = \text{min}(T_i, C_i) I(\text{min}(T_i, C_i) + B_i \leq t).
\end{equation*}
This can also be applied to our set-up. It is clear that the chart will not have a continuous drift with this specification, but only jumps in value whenever an observation has experienced an event or censoring. 

Another circumstance that could appear is when the information is only available at certain time points. An example is when a patient is diagnosed in the middle of a year, but the observation is only included in the database at the end of the year. A possible solution is to again redefine $A_i(t)$ so that the observation is included in the system periodically, e.g. at the start of every new year so that 

\begin{equation*}
     A_i(t) =\min(T_i,C_i,t-B_i) I(\lceil B_i \rceil \leq t),
\end{equation*}
where $\lceil \cdot \rceil$ is the ceiling function. 

Similar to the above, we can extend the first situation where the information of an observation is only available periodically after the observation experiences an event or has been censored, e.g. when a patient either dies or drops out of a study in the middle of a year and the database is updated with the given observation at the start of the new year. Then, $A_i(t)$ can be defined as follows: 

\begin{equation*}
    A_i(t) = \text{min}(T_i, C_i) I(\lceil \text{min}(T_i, C_i) + B_i \rceil \leq t).
\end{equation*}

Another scenario is when the entire database is updated in regular intervals, adding all information accumulating since the last update. Then, an option would be to run the CUSUM retrospectively by running the CUSUM for the last period once the data for that period become available. This would of course lead to a certain delay in signalling versus when running the chart in real-time. 

In some cases, it is of interest to just monitor for survival up to a certain time after diagnosis $t_D$, for instance $t_D=5$ years survival in cancer patients.  A benefit with such monitoring is that we only need baseline data over a time period long enough to be able to estimate $h_{E,0}(\cdot)$ up to $t_D$. Furthermore, with this set-up, there is no restriction as to how long we can run the monitoring as we will only need to evaluate $h_{E,0}(t)$ for $t\leq t_D$. Thus, it is not required to have an estimate of $h_{E,0}(t)$ for $t\geq t_D$. Estimation of the in-control model is further discussed in the next subsection.

\subsection{Estimating the in-control model}
\label{subsec:estincontrol} 

In practical applications, the true in-control excess hazard will be unknown and has to be estimated from baseline data.
To run the monitoring procedure, we need to be able to evaluate $h_{E,0}(t)$ and $H_{E,0}(t)$ for any survival time $t$ that could be observed during the monitoring. In practice, this means that if nonparametric estimates of $h_{E,0}(t)$ and $H_{E,0}(t)$ are used, baseline data for at least as long period as the maximum monitoring period would be required to estimate these functions, if the monitoring is run without any upper limit on the time to event of interest. If a parametric estimate is used, one could in principle use the estimated model beyond the event horizon observed in the baseline data.

When estimating the in-control excess hazard, the estimation error will propagate to the achieved in-control performance. The impact of this, as well as the impact of model misspecifications, will be illustrated in Section \ref{subsec:estimation_error}.

\section{Simulation study}
\label{sec:sim} 
In this section, properties of the suggested procedures are studied by simulations.

\subsection{Simulation set-up}
\label{subsec:sim:set-up}

For the following, we will use a proportional excess hazard model, i.e. $h_{E,0}(\cdot,X_i) = h_0(\cdot) \exp(\mathbf \beta X_i)$. Inspired by data from the Norwegian Cancer Registry,  Table \ref{table:covariates_sim} 
in Appendix~\ref{appendix:summary_table} presents the covariates that we will consider in the simulations, with corresponding parameter values.  

The interim censoring time is set to be exponentially distributed with the rate parameter equal to 0.000275.  The monitoring is run for 10 years (in one case 5 years) and is using the population hazard of Norway from the beginning of 2010 to the end of 2019. The baseline excess hazard is chosen to be a piecewise constant function of the form $h_0(t)=\exp{\squarebrackets{\sum_{k} \chi_k I_k(t)}}$, where $I_k(t)=1$ whenever $t$ lies in the $k$-th band of the 10-year follow-up interval. Here, we partition the follow-up interval into yearly bands during the first five years before defining one single band for $t \in [5, 10]$. Inspired by the cancer data, we will use $\chi = (-1.4, -1.6, -1.8, -2.0, -2.1, -3.0)$. 
		
		

In most simulations, we assume that both the distributions of the covariates and the parameter vectors $\beta$ and $\chi$ are known. Thus, the charts are calculated using the true parameters, and hence the true functional form of $h_0$. However,  the impact of estimation error will be studied in Section~\ref{subsec:estimation_error}. 

 The arrival of patients is simulated as a homogeneous Poisson process with an arrival rate $\lambda_a$. To determine the threshold $c$, we use the false signal probability during the 10 years (in one case 5 years) of monitoring as criteria. The threshold is then computed by simulating 1000 or $10\,000$ CUSUMs from the in-control model, and we tune the threshold to achieve the desired false signal probability, which we specify to either 0.05 or 0.01. After determining the signal threshold $c$, the signal probabilities for various out-of-control situations are simulated by simulating 1000 or $10\,000$ CUSUMs, the former being used for larger values of $\lambda_a$ to obtain the results in a reasonable time. 

We study three different scenarios for the shift from in-control to out-of-control: i) All patients are in the out-of-control situation from the beginning ($\eta=0$), ii) All patients change to the out-of-control hazard at some time point $\eta$ years after monitoring start (either $\eta=2.5$ or $\eta=5$), iii) Only patients arriving after some time point $\eta^*$ have the out-of-control hazard (either $\eta^*=2.5$ or $\eta^*=5$). In all cases, the data are simulated from the correct specification of the CUSUM chart, i.e. the CUSUM charts are correctly specified for the incoming data. The reason for choosing two different change-point times and monitoring lengths is due to the differences in sample sizes simulated, resulting in differences in how fast a signal is obtained.

\subsection{Proportional alternative}
\label{subsec:sim:proportional}
This subsection considers different scenarios with the proportional alternative represented by \eqref{eq:loglik_prop}.  The charts are computed using the correct value of $\rho$ that determines the various out-of-control situations. 
The first simulation set-up examines the scenario of four different values of $\rho$ in a situation with a moderate yearly arrival rate of $\lambda_a = 250$. The true change-point occurs 5 years after the monitoring start.

 The results from the described simulation set-up are given in Table \ref{table:signal_results_prop_nr2}. One can observe that in the situation when the excess hazard is in the out-of-control state from the start, almost all simulations yield a signal during the 10-year monitoring period for the largest shifts, i.e. $\rho = 0.80$ and $\rho = 1.20$. When the change is less pronounced like $\rho = 0.90$ and $\rho = 1.10$, the signal ratio is reduced, and of course, lower for the 0.01 false signal probability versus 0.05. If the change occurs later in the monitoring period, it is natural that the signal ratio is further decreased as there is less time for the charts to signal, hence the lower signal ratios for situations with $\eta = 5$. Finally, the situation where only the individuals arriving after 5 years will experience the out-of-control state yields the smallest signal ratio. As expected, the proposed CUSUM charts have less  power to detect these shifts. 

\begin{table}[htbp]
\centering
\begin{tabular}{ccccc}
\toprule
 &  & \multicolumn{3}{c}{Signal Ratio} \\
 \cmidrule(lr){3-5}
 & Scenario & $\eta = 0$ & $\eta = 5$ & $\eta^* = 5$ \\
 \cmidrule(lr){2 - 2}
$\rho$ & $P(\tau \leq 10 \mid \eta = \infty)$
 &  &  &  \\
\midrule
\multirow{2}{*}{0.80} & 0.01 ($c = 6.41$) & 95.34\% & 78.39\% & 42.97\% \\
 & 0.05 ($c = 5.02$) & 98.50\% & 89.85\% & 64.17\% \\
\cline{1-5}
\multirow{2}{*}{0.90} & 0.01 ($c = 4.37$) & 43.57\% & 21.49\% & 9.21\% \\
 & 0.05 ($c = 3.41$) & 65.28\% & 42.61\% & 24.44\% \\
\cline{1-5}
\multirow{2}{*}{1.10} & 0.01 ($c = 4.32$) & 36.09\% & 17.39\% & 8.02\% \\
 & 0.05 ($c = 3.36$) & 59.78\% & 37.78\% & 21.40\% \\
\cline{1-5}
\multirow{2}{*}{1.20} & 0.01 ($c = 6.12$) & 91.20\% & 67.78\% & 32.45\% \\
 & 0.05 ($c = 4.80$) & 96.75\% & 83.95\% & 54.09\% \\
\cline{1-5}
\bottomrule
\end{tabular}%
\caption{The table reports proportions of cases when the CUSUM signals in a setting where individuals arrive over calendar time according to a Poisson process with rate $\lambda_a=250$, and the out-of-control model is $h_{1i}(\cdot)=h_P(\cdot,Z_i)+\rho h_{E,0}(\cdot,X_i)$ with the latter having a piecewise constant baseline excess hazard. Different shifts in terms of $\rho$ and different shift points $\eta$ are considered. For the shift after 5 years of monitoring, both the setting that the shift affects all individuals ($\eta = 5$) and the setting when the shift only affects new individuals ($\eta^* = 5$)  are considered. $10\,000$ simulations are performed for each combination of $\rho$ and scenario of shift, and for finding thresholds $c$.}
\label{table:signal_results_prop_nr2}
\end{table}


\begin{table}[htbp]
\centering
\begin{tabular}{ccccc}
\toprule
 &  & \multicolumn{3}{c}{Signal Ratio} \\
 \cmidrule(lr){3-5}
 & Scenario & $\eta = 0$ & $\eta = 2.5$ & $\eta^* = 2.5$ \\
 \cmidrule(lr){2 - 2}
$\rho$ & $P(\tau \leq 5 \mid \eta = \infty)$
 &  &  &  \\
\midrule
\multirow{2}{*}{0.90} & 0.01 ($c = 8.00$) & 97.3\% & 84.9\% & 26.0\% \\
 & 0.05 ($c = 6.23$) & 99.2\% & 94.0\% & 46.0\% \\
\cline{1-5}
\multirow{2}{*}{0.95} & 0.01 ($c = 6.06$) & 37.2\% & 16.4\% & 3.0\% \\
 & 0.05 ($c = 4.96$) & 54.8\% & 33.3\% & 8.5\% \\
\cline{1-5}
\multirow{2}{*}{1.05} & 0.01 ($c = 6.08$) & 33.3\% & 13.8\% & 3.3\% \\
 & 0.05 ($c = 4.69$) & 57.6\% & 37.1\% & 10.1\% \\
\cline{1-5}
\multirow{2}{*}{1.10} & 0.01 ($c = 7.63$) & 96.0\% & 84.6\% & 22.2\% \\
 & 0.05 ($c = 6.06$) & 98.5\% & 92.5\% & 42.0\% \\
\cline{1-5}
\bottomrule
\end{tabular}%
\caption{The table reports proportions of cases when the CUSUM signals in a setting where individuals arrive over calendar time according to a Poisson process with rate $\lambda_a=3700$, and the out-of-control model is $h_{1i}(\cdot)=h_P(\cdot,Z_i)+\rho h_{E,0}(\cdot,X_i)$ with the latter having a piecewise constant baseline excess hazard. The CUSUM is run for 5 years. Different shifts in terms of $\rho$ and different shift points $\eta$ are considered. For the shift after 2.5 years of monitoring, both the setting that the shift affects all individuals ($\eta = 2.5$) and the setting when the shift only affects new individuals ($\eta^* = 2.5$)  are considered. 1000 simulations are performed for each combination of $\rho$ and scenario of shift, and for finding thresholds $c$.}
\label{table:signal_results_prop_nr3}
\end{table}

Next, a situation with a much larger arrival rate, similar to what is observed in the cancer registry data used in Section~\ref{sec:realdata}, is considered by letting $\lambda_a = 3700$. If one considers the same change-point time of 5 years after monitoring start and a monitoring period of 10 year, this situation will lead to very large detection ratios close to 100\% in many cases. This is naturally intuitive as the power will increase with larger sample sizes. To balance out the increased power, we just run the CUSUM for 5 years and set an earlier change-point of 2.5 years after the start of monitoring and examine less prominent changes in the excess hazard by investigating values of $\rho$ that differ less from unity compared to the previous situation. From Table \ref{table:signal_results_prop_nr3}, the power of the monitoring system is relatively high if the excess hazard changes by 10\% for both scenarios of $\eta$. On the other hand, even with the much larger amount of arrivals, just above half of the charts will signal when the false signal probability is $0.05$ and the excess hazard changes by 5\%. 

\subsection{Additive alternative}

\begin{table}[htbp]
\centering
\begin{tabular}{ccccc}
\toprule
 &  & \multicolumn{3}{c}{Signal Ratio} \\
 \cmidrule(lr){3-5}
 & Scenario & $\eta = 0$ & $\eta = 5$ & $\eta^* = 5$ \\
 \cmidrule(lr){2 - 2}
$\gamma$ & $P(\tau \leq 10 \mid \eta = \infty)$
 &  &  &  \\
\midrule
\multirow{2}{*}{-0.002} & 0.01 ($c = 4.93$) & 61.8\% & 49.3\% & 5.0\% \\
 & 0.05 ($c = 3.78$)& 79.3\% & 70.7\% & 17.6\% \\
\cline{1-5}
\multirow{2}{*}{0.002} & 0.01 ($c = 4.78$) & 60.6\% & 46.9\% & 5.6\% \\
 & 0.05 ($c = 3.83$)& 76.6\% & 64.7\% & 13.5\% \\
\cline{1-5}
\multirow{2}{*}{0.005} & 0.01 ($c = 6.81$) & 100.0\% & 100.0\% & 22.5\% \\
 & 0.05 ($c = 5.50$) & 100.0\% & 100.0\% & 40.0\% \\
\cline{1-5}
\bottomrule
\end{tabular}%
\caption{The table reports proportions of cases when the CUSUM signals in a setting where individuals arrive over calendar time according to a Poisson process with rate $\lambda_a=3700$, and the out-of-control model is $h_{1i}(\cdot)=h_P(\cdot,Z_i)+\text{max}(0, h_{E,0}(\cdot,X_i) + \gamma)$ with the latter having a piecewise constant baseline excess hazard. Different shifts in terms of $\gamma$ and different shift points $\eta$ are considered. For the shift after 5 years of monitoring, both the setting that the shift affects all individuals ($\eta = 5$) and the setting when the shift only affects new individuals ($\eta^* = 5$) are considered. 1000 simulations are performed for each combination of $\gamma$ and scenario of shift, and for finding thresholds $c$.}
\label{table:signal_results_add_nr1}
\end{table}

In this subsection, we perform a similar experiment as in the preceding for the additive alternative given by \eqref{eq:loglik_add}. By letting $\lambda_a=3700$, we examine three values of $ \gamma$: -0.002, 0.002 and 0.005. It turns out that combining the general simulation set-up described earlier with $\gamma = -0.002$ yields individual out-of-control excess hazards that are non-negative so that \eqref{eq:loglik_add_without_max} can also be used for faster calculations. For smaller shifts like $0.002$ in absolute value, we can see from Table \ref{table:signal_results_add_nr1} that the power of the charts is moderate when using the threshold giving a 5\% probability of false signal. On the other hand, increasing the effect of change to $0.005$ leads to signals for all simulated charts for the cases where all individuals are affected by the shift. We also notice that the $\gamma=-0.002$ and $\gamma=0.002$ cases achieve roughly the same power. This is similar to the case with proportional alternative, if the CUSUM chart is set to monitor a large change and this is indeed the case in reality, the chart will detect and signal much faster compared to the situation with less pronounced change that is harder to identify. As before, the situation where only new arrivals are affected by the change has the smallest signal ratios for all values of $\gamma$. 

\subsection{Linear accelerated time alternative}

\begin{table}[htbp]
\centering
\begin{tabular}{ccccc}
\toprule
 &  & \multicolumn{3}{c}{Signal Ratio} \\
 \cmidrule(lr){3-5}
 & Scenario & $\eta = 0$ & $\eta = 5$ & $\eta^* = 5$ \\
 \cmidrule(lr){2 - 2}
$k$ & $P(\tau \leq 10 \mid \eta = \infty)$
 &  &  &  \\
\midrule
\multirow{2}{*}{0.90} & 0.01 ($c = 7.83$) & 100.0\% & 100.0\% & 44.7\% \\
 & 0.05 ($c = 6.50$) & 100.0\% & 100.0\% & 62.8\% \\
\cline{1-5}
\multirow{2}{*}{0.95} & 0.01 ($c = 7.29$) & 99.9\% & 99.7\% & 10.6\% \\
 & 0.05 ($c = 5.89$) & 99.9\% & 100.0\% & 25.4\% \\
\cline{1-5}
\multirow{2}{*}{1.05} & 0.01 ($c = 7.18$) & 99.9\% & 99.6\% & 14.3\% \\
 & 0.05 ($c = 5.97$)& 100.0\% & 99.9\% & 26.2\% \\
\cline{1-5}
\multirow{2}{*}{1.10} & 0.01 ($c = 8.49$) & 100.0\% & 100.0\% & 35.3\% \\
 & 0.05 ($c = 6.47$) & 100.0\% & 100.0\% & 64.2\% \\
\cline{1-5}
\bottomrule
\end{tabular}%
\caption{The table reports proportions of cases when the CUSUM signals in a setting where individuals arrive over calendar time according to a Poisson process with rate $\lambda_a=3700$, and the out-of-control model is $h_{1i}(\cdot)=h_P(\cdot,Z_i)+k h_{E,0}(k \cdot,X_i)$ with the latter having a piecewise constant baseline excess hazard. Different shifts in terms of $k$ and different shift points $\eta$ are considered. For the shift after 5 years of monitoring, both the setting that the shift affects all individuals ($\eta = 5$) and the setting when the shift only affects new individuals ($\eta^* = 5$) are considered. 1000 simulations are performed for each combination of $k$ and scenario of shift, and for finding thresholds $c$.}
\label{table:signal_results_acc_lin_time_nr1}
\end{table}

For the last set-up in this section, we investigate the performance of the linear accelerated time alternative (\ref{eq:acc_lin_time}) with four different values of $k$. When $k > 1$, we extend the upper limit of the last band of $h_0(t)=\exp{\squarebrackets{\sum_{k} \chi_k I_k(t)}}$ covering the time interval between 5 and 10 years to $10 k$ years. Table \ref{table:signal_results_acc_lin_time_nr1} shows that almost all charts signal for the situations when $\eta = 0$ and $\eta = 5$ across all four values of $k$. For $\eta^* = 5$, we can still observe that the procedure struggles more to capture these changes.

\subsection{Estimation error and model misspecification}
\label{subsec:estimation_error}

In the preceding simulation studies, the true covariate parameters and functional form of the baseline excess hazard are used when computing the CUSUM charts. In practice, these are all unknown quantities and need to be estimated as discussed in Section \ref{subsec:estincontrol}. To illustrate the effect of estimation error in achieving e.g. the specified in-control false probability or average run length, we perform the simulation set-up presented in Algorithm \ref{alg:simulation_set-up_estimation_error}.


\begin{algorithm}
    \caption{Simulation set-up to illustrate effect of estimation error and model misspecification}
    \begin{algorithmic}[1]
    \STATE Define a specific underlying proportional excess hazard model $h_{E,0}(\cdot, X_i)=h_0(\cdot)\exp(\beta X_i)$ with $h_0$ being a parametric baseline excess hazard function. 
    \STATE From the true model, simulate a data set to be used for estimating the in-control model. 
    \STATE Based on the simulated data set, estimate an in-control model (which could be an estimated "true" model, e.g. a model with a piecewise constant baseline if the true excess hazard is on this form, or an estimated "wrong model", e.g. with a smooth baseline if the true excess hazard is piecewise constant.) 
    \STATE Treat the estimated model as a true model and simulate the threshold $c$ to obtain a certain false signal probability under this assumed true model by simulating observations from the in-control model. Two possibilities regarding the covariate distribution will be considered here: In the first approach, the true covariate distribution  is assumed to be known, the second uses a nonparametric bootstrap procedure on the simulated data set from step 2. 
    \STATE Use the estimated model from step 3 with the threshold from step 4 and run the CUSUM many times on data simulated from the true model specified in step 1. The achieved signal probability is calculated using these runs.
    \STATE Repeat step 2 to step 5 many times in order to characterize the distribution of the in-control false signal probability obtained under the estimated model.  
    \end{algorithmic}
    \label{alg:simulation_set-up_estimation_error}
\end{algorithm}

Furthermore, we will also consider some model misspecification aspects. One would presume that the observed in-control false signal probability will be further off from the nominal value if the specified form of the baseline is not correct, e.g. when using a piecewise constant baseline excess hazard model when the true excess hazard is a continuous function. Consequently, we will consider two different types of baseline excess hazard in this subsection: A piecewise constant baseline and a flexible smooth nonparametric baseline hazard. Both methods can be found in the R package \texttt{relsurv} via the \texttt{rsadd}-function, see e.g. \citeN{perme2018nonparametric}. A note regarding the smoothing procedure to obtain continuous nonparametric estimates of the baseline excess hazard is given in Appendix \ref{appendix:smoothing}.

In addition, we also investigate for each combination of estimation procedure and true excess hazard if the results obtained from bootstrapping the covariate distributions from the data set obtained in step 2 of Algorithm \ref{alg:simulation_set-up_estimation_error} differ from simulating directly from the true covariate distributions. We will only perform the simulation using the proportional alternative, but the idea is the same for the remaining out-of-control situations. 

\subsubsection{Piecewise constant baseline}
\begin{figure}
     \centering
     \begin{subfigure}[t]{0.49\textwidth}
         \centering
         \includegraphics[width=\textwidth]{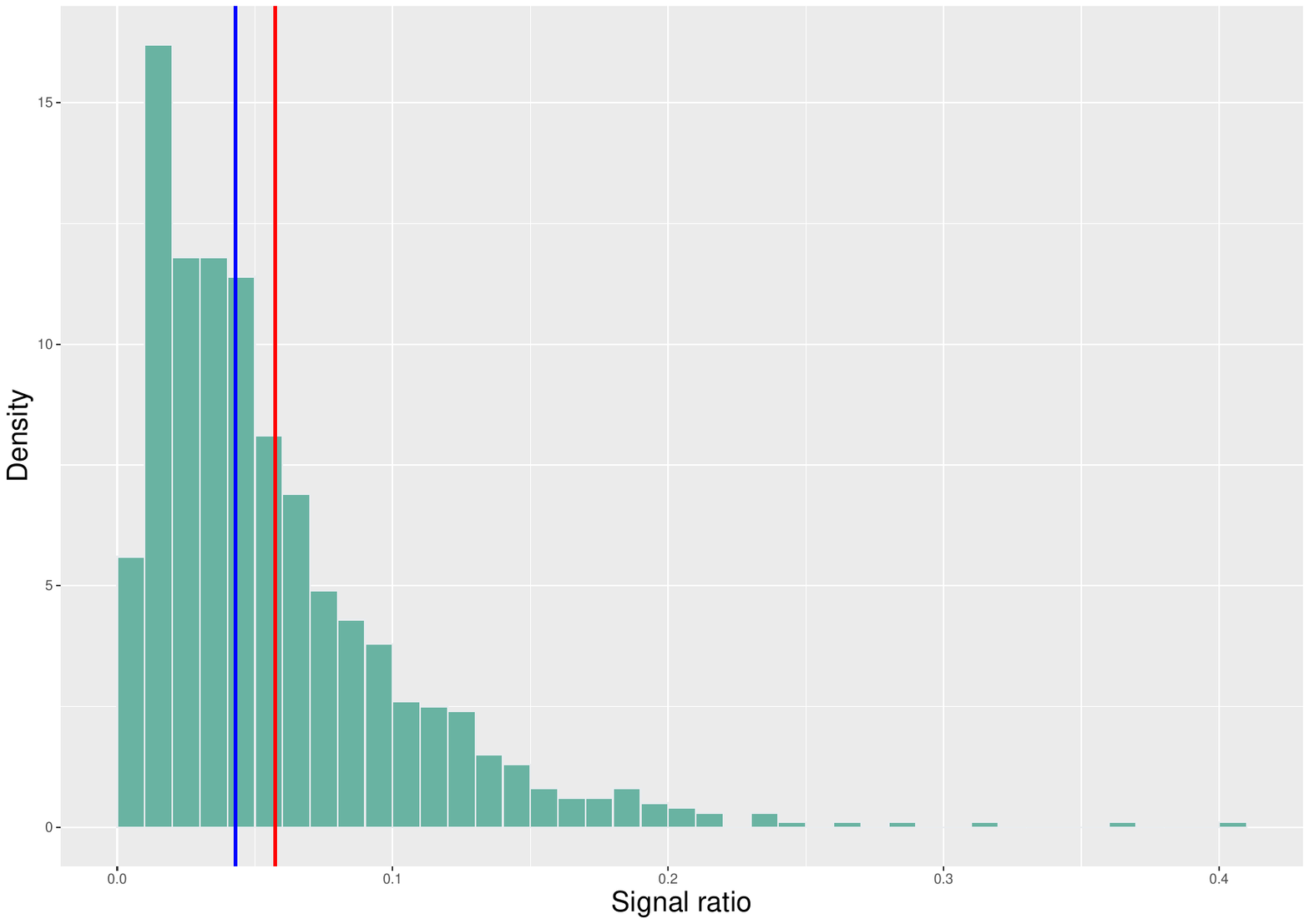}
         \caption{True covariate distribution used in step 4. Mean signal ratio is 5.73\%, median signal ratio is 4.30\%.}
         \label{fig:in_control_piecewise_piecewise_nr1}
     \end{subfigure}
     \hfill
     \begin{subfigure}[t]{0.49\textwidth}
         \centering
         \includegraphics[width=\textwidth]{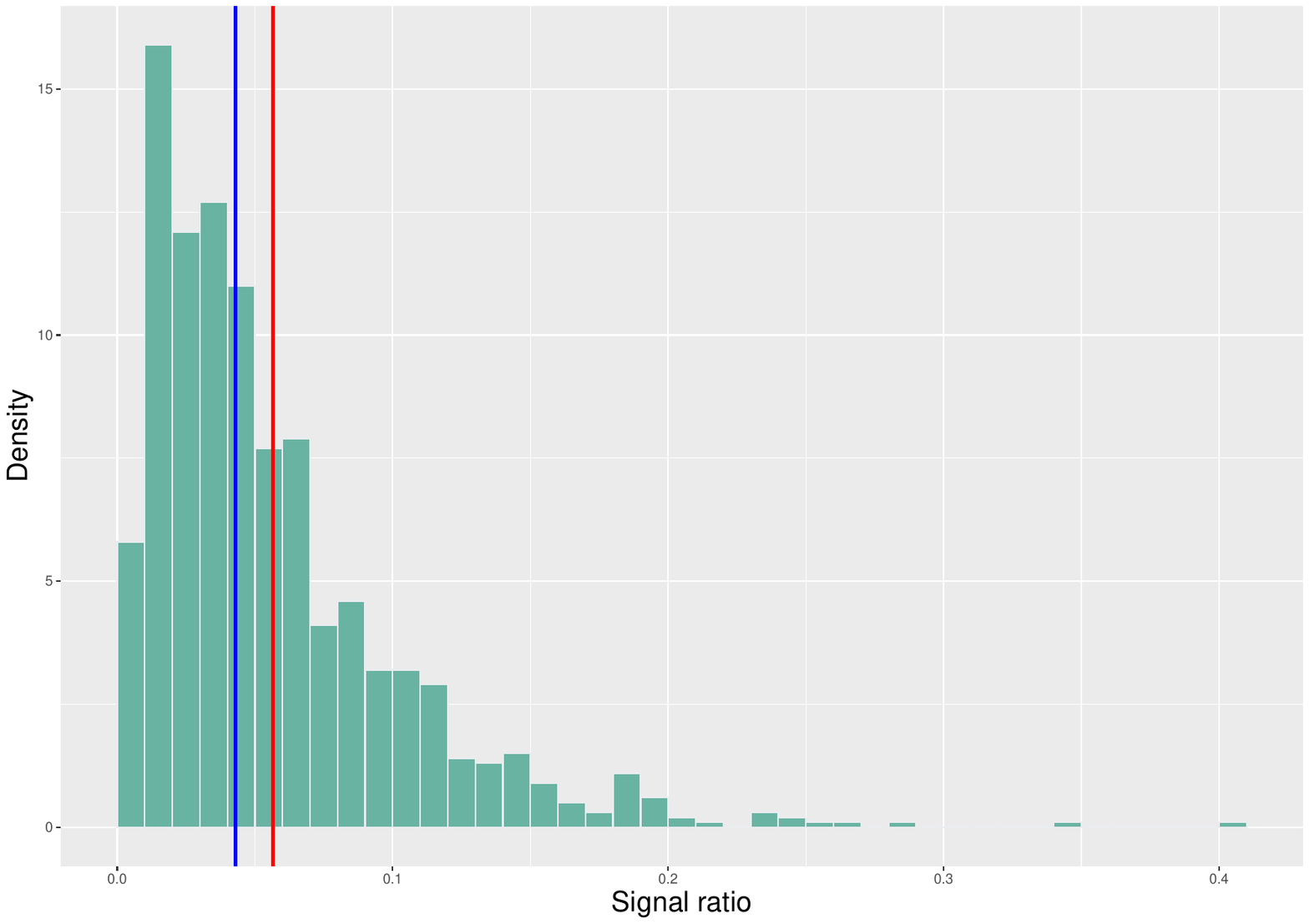}
         \caption{Bootstrapping used to estimate the covariate distribution. Mean signal ratio is 5.65\%, median signal ratio is 4.30\%.}
         \label{fig:in_control_piecewise_piecewise_nr2}
     \end{subfigure}
     \caption{Histograms of the achieved signal probabilities under in-control scenario with the true baseline excess hazard following a piecewise constant function. The model estimation is done using the model with piecewise constant baseline in step 3. Here, the blue vertical line represents the median and the red line corresponds to the mean signal ratio.}
     \label{fig:in_control_piecewise_piecewise}
\end{figure}

In this part of the simulation, we consider the same piecewise constant baseline excess hazard set-up presented previously in this section. The arrival rate is set to $\lambda_a = 3750$, the proportionality constant relating the out-of-control and in-control scenario is $\rho = 0.90$. We let the desired in-control false probability during the first five years of monitoring be equal to $P(\tau \leq 5 \mid \eta = \infty) = 0.05$. Steps 2 to 5 are performed 1000 times to obtain the distribution of in-control false probability under a given combination of estimation method and covariate simulation. For each iteration, 1000 simulations are also done in order to obtain the threshold value $c$. Figure \ref{fig:in_control_piecewise_piecewise} shows the results when the model estimation is performed using the piecewise baseline excess hazard model. 
This resembles the situation where the estimated baseline excess hazard is correctly specified. We can observe that both the median and mean in-control signal ratio, regardless if the covariate distributions are simulated using bootstrap or the true distributions, only differ slightly from the desired nominal value of 5\%  due to the effect of the estimation error of $\beta$ and $\chi$.  



On the other hand, when examining the results of the same procedure using the smooth baseline excess hazard estimate, we see that the median and mean in-control signal ratio, independent of the scenario considered in step 4, are  further away from the nominal value. Not only do we have the estimation error in the covariate parameters that affects the results, the incorrect specification of the form of the baseline excess hazard when using the smooth baseline estimate leads to larger deviation from the desired false signal ratio. Therefore, if the true baseline excess hazard is indeed a piecewise constant function, approximating it with a continuous and smooth function might cause undesired behaviour in the monitoring procedure. 

\begin{figure}
     \centering
     \begin{subfigure}[t]{0.49\textwidth}
         \centering
         \includegraphics[width=\textwidth]{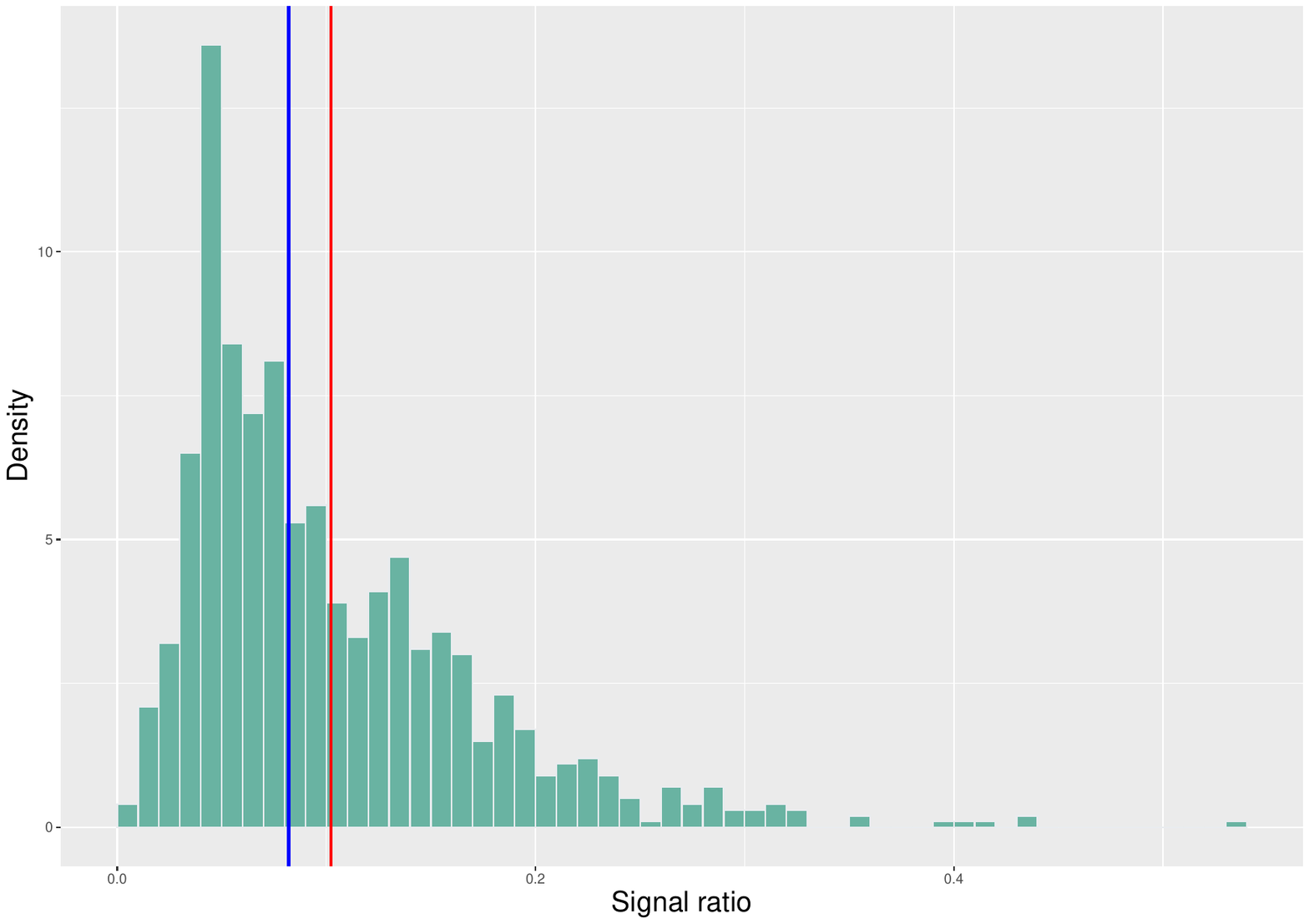}
         \caption{True covariate distribution used in step 4. Mean signal ratio is 10.23\%, median signal ratio is 8.20\%.}
         \label{fig:in_control_piecewise_em_nr1}
     \end{subfigure}
     \hfill
     \begin{subfigure}[t]{0.49\textwidth}
         \centering
         \includegraphics[width=\textwidth]{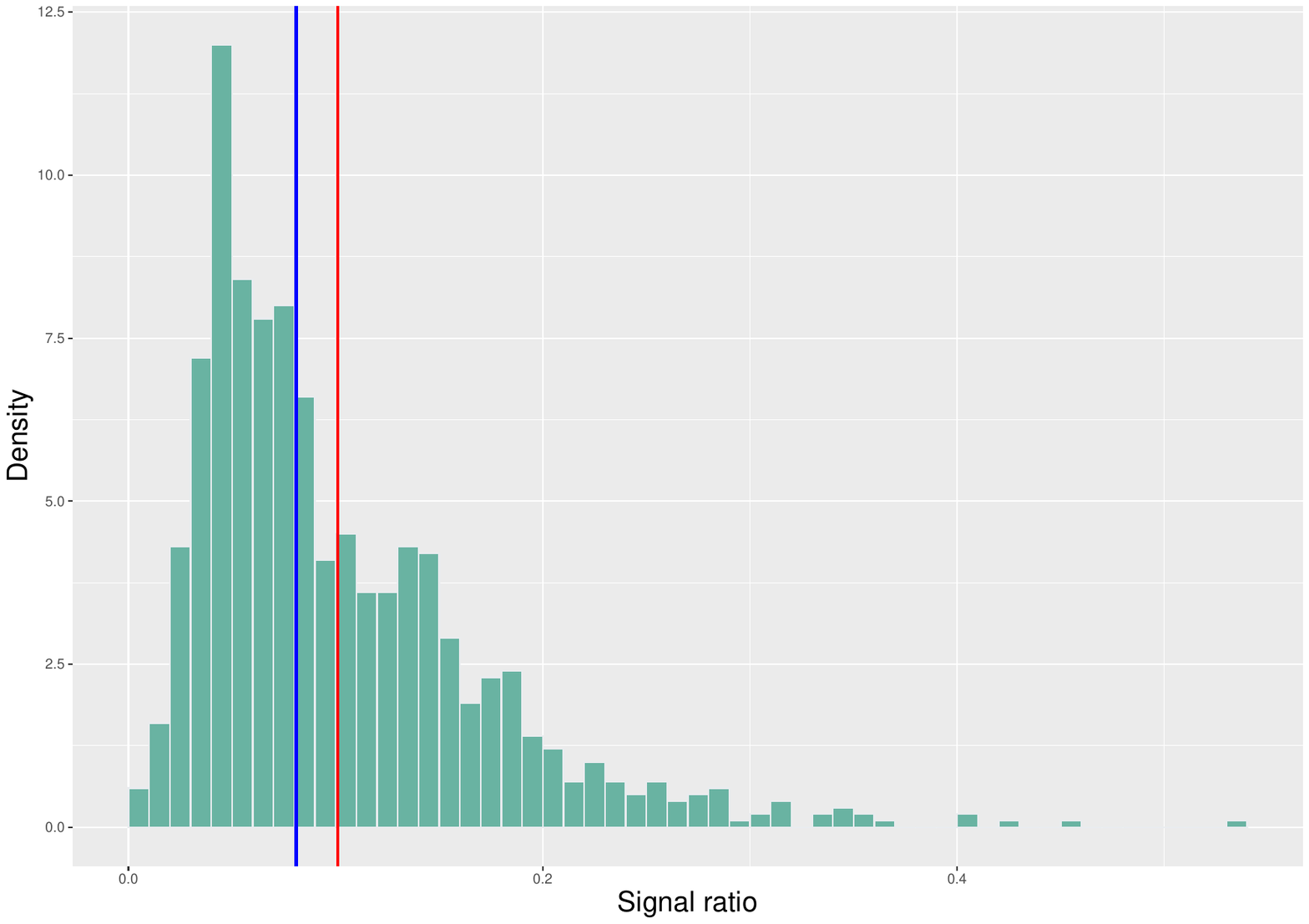}
         \caption{Bootstrapping used to estimate the covariate distribution. Mean signal ratio is 10.10\%, median signal ratio is 8.10\%.}
         \label{fig:in_control_piecewise_em_nr2}
     \end{subfigure}
     \caption{Histograms of the achieved signal probabilities under in-control scenario with the true baseline excess hazard following a piecewise constant function. The model estimation is done using a smooth nonparametric baseline excess hazard in step 3. Here, the blue vertical line represents the median and the red line corresponds to the mean signal ratio.}
     \label{fig:in_control_piecewise_em}
\end{figure}

\subsubsection{Weibull baseline}
\begin{figure}
     \centering
     \begin{subfigure}[t]{0.49\textwidth}
         \centering
         \includegraphics[width=\textwidth]{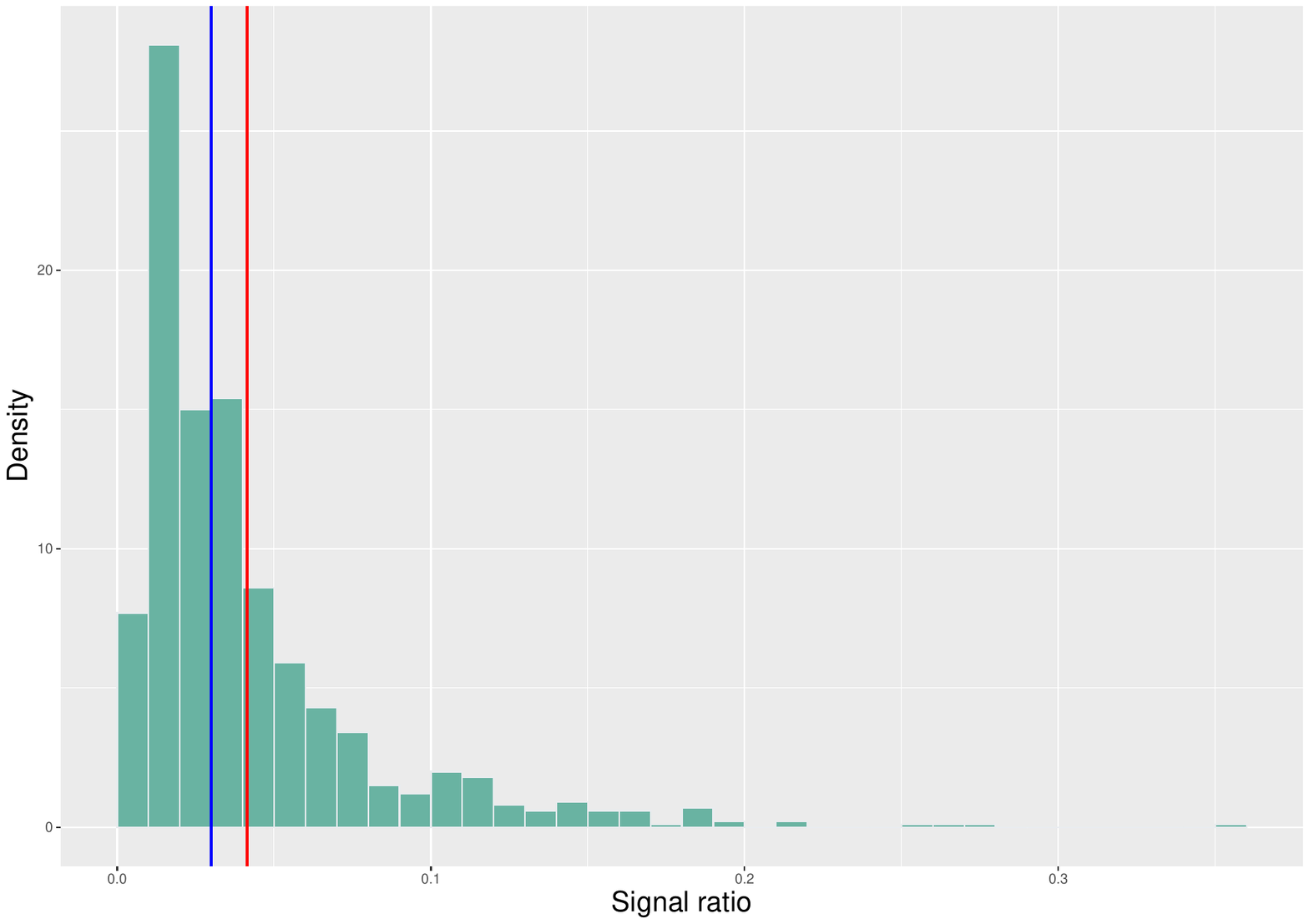}
         \caption{True covariate distribution used in step 4. Mean signal ratio is 4.14\%, median signal ratio is 3.00\%.}
         \label{fig:in_control_weibull_piecewise_nr1}
     \end{subfigure}
     \hfill
     \begin{subfigure}[t]{0.49\textwidth}
         \centering
         \includegraphics[width=\textwidth]{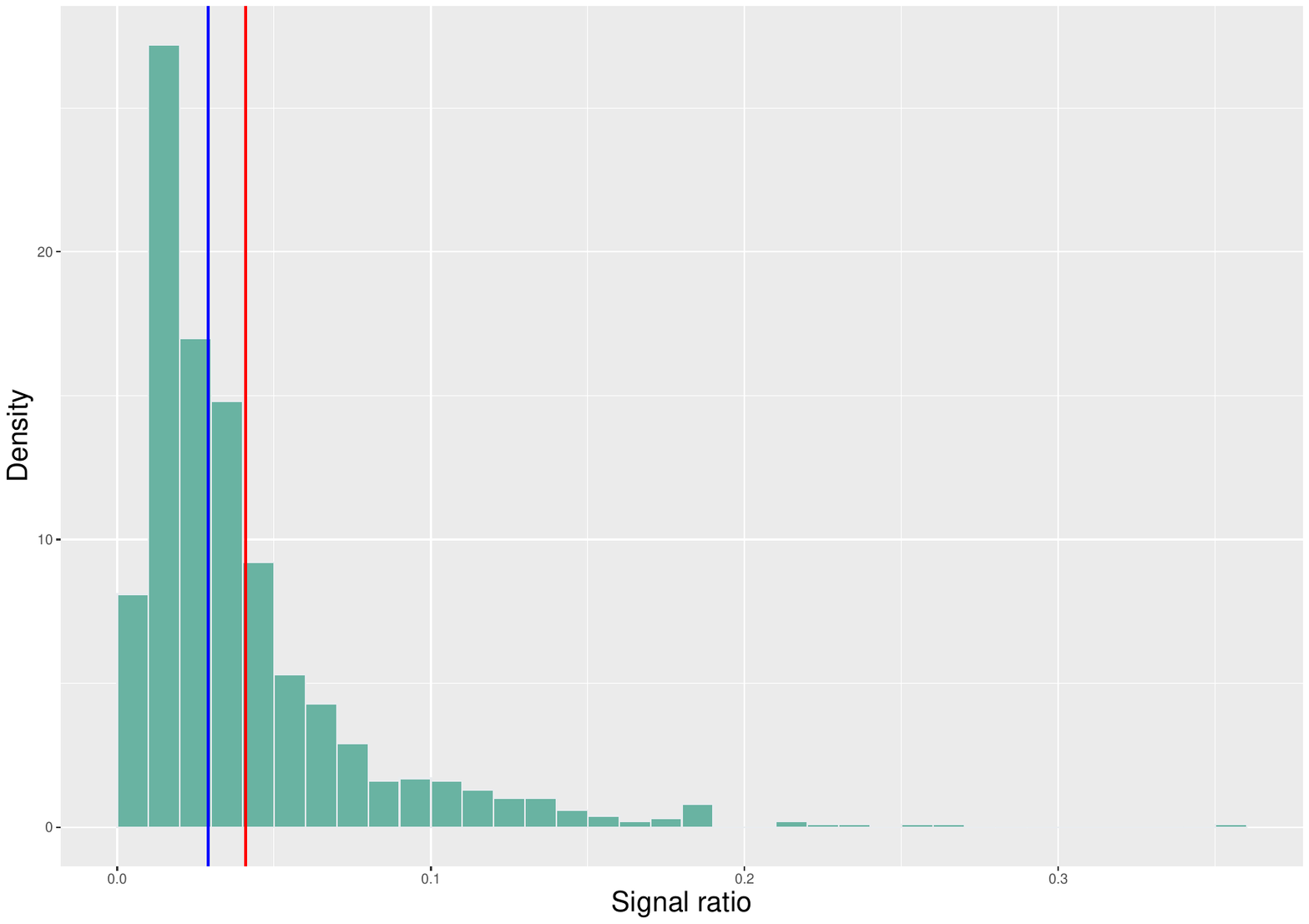}
         \caption{Bootstrapping used to estimate the covariate distribution. Mean signal ratio is 4.09\%, median signal ratio is 2.90\%.}
         \label{fig:in_control_weibull_piecewise_nr2}
     \end{subfigure}
     \caption{Histograms of the achieved signal probabilities under in-control scenario with the true baseline excess hazard following the hazard function of the Weibull distribution. The model estimation is done using the model with piecewise constant baseline in step 3. Here, the blue vertical line represents the median and the red line corresponds to the mean signal ratio.}
     \label{fig:in_control_weibull_piecewise}
\end{figure}

In theory, the previous true form of baseline excess hazard favors the model with piecewise constant baseline as it is difficult for the estimated smoothed nonparametric baseline excess hazard to capture the stepwise behaviour. We now illustrate the opposite scenario by considering a Weibull baseline as the true form, where 

\begin{equation*}
    h_0(t) = a b t ^{a - 1}.
\end{equation*}
 Inspired by the data from the Norwegian Cancer Registry, we let $a = 0.65$ and $b = 0.25$. Otherwise, the other quantities remain the same as in the piecewise constant baseline scenario. Results are reported in Figure~\ref{fig:in_control_weibull_piecewise}. Compared to the situation with a piecewise constant baseline as the true form, the differences between the 5\% nominal value and the mean and median signal ratio are now increased as expected due to the misspecification of the baseline excess hazard in the model estimation. 

\begin{figure}
     \centering
     \begin{subfigure}[t]{0.49\textwidth}
         \centering
         \includegraphics[width=\textwidth]{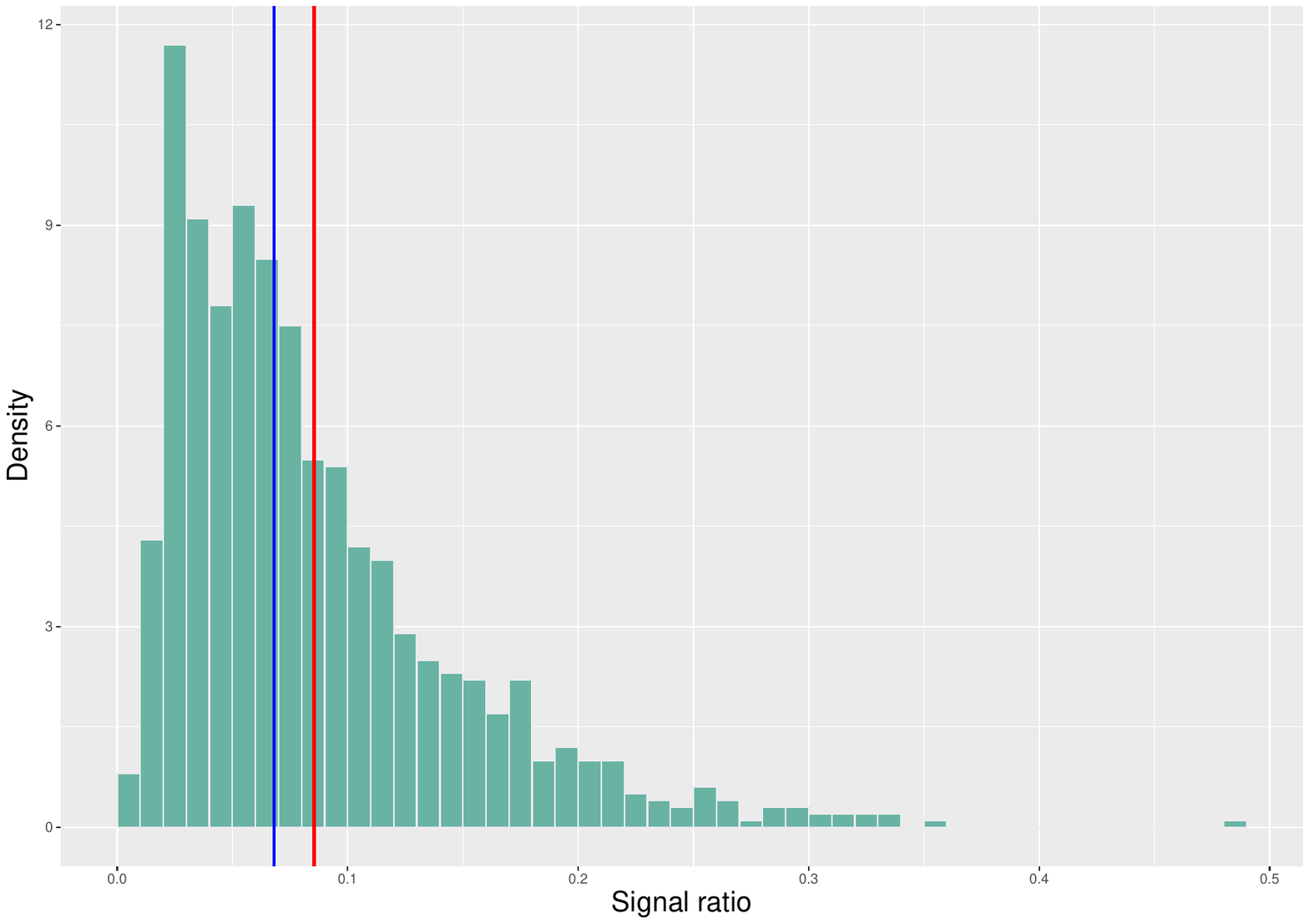}
         \caption{True covariate distribution used in step 4. Mean signal ratio is 8.55\%, median signal ratio is 6.80\%.}
         \label{fig:in_control_weibull_em_nr1}
     \end{subfigure}
     \hfill
     \begin{subfigure}[t]{0.49\textwidth}
         \centering
         \includegraphics[width=\textwidth]{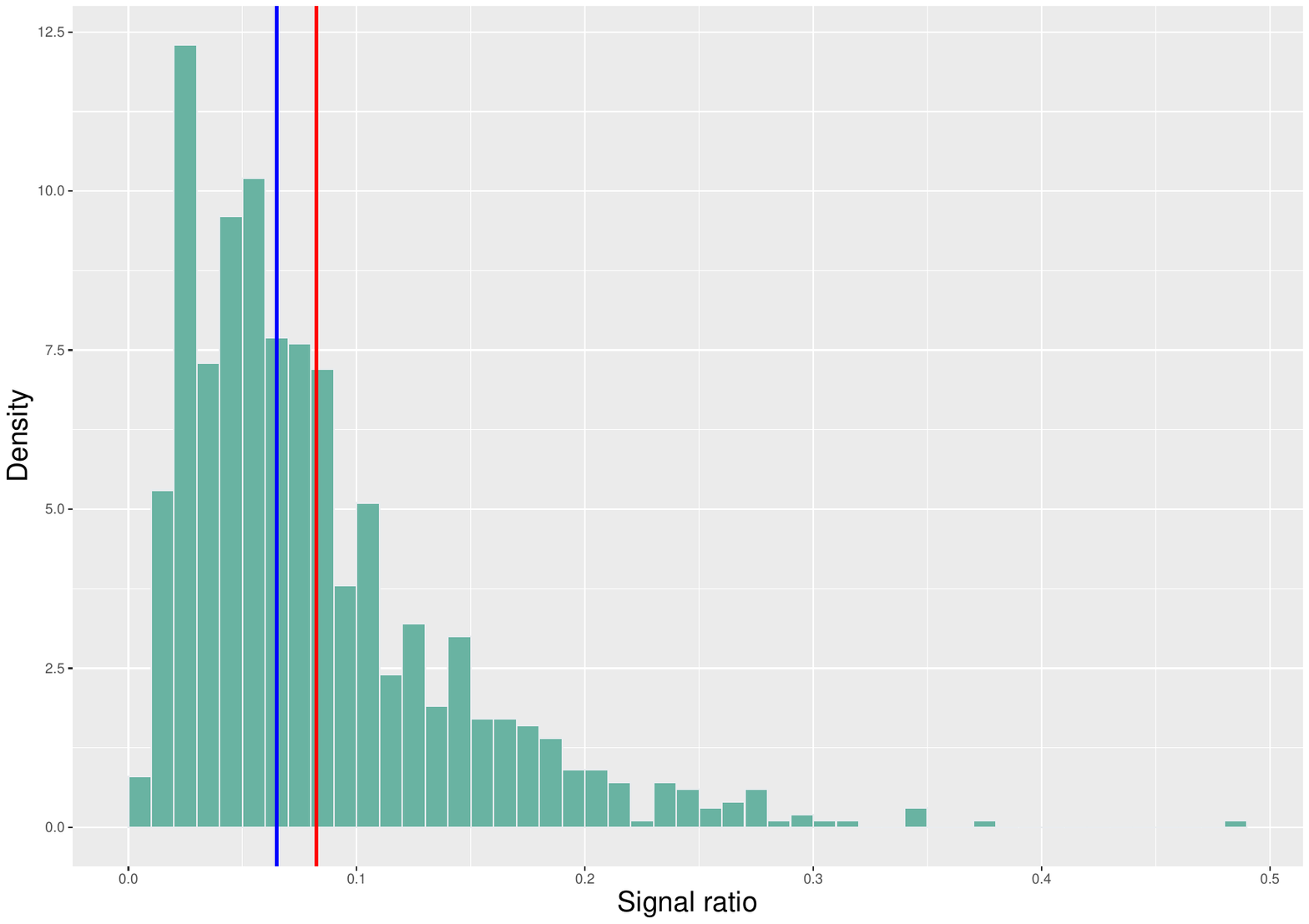}
         \caption{Bootstrapping used to estimate the covariate distribution. Mean signal ratio is 8.24\%, median signal ratio is 6.50\%.}
         \label{fig:in_control_weibull_em_nr2}
     \end{subfigure}
     \caption{Histograms of the achieved signal probabilities under in-control scenario with the true baseline excess hazard following the hazard function of the Weibull distribution. The model estimation is done using a smooth nonparametric baseline excess hazard. Here, the blue vertical line represents the median and the red line corresponds to the mean signal ratio.}
     \label{fig:in_control_weibull_em}
\end{figure}

Figure~\ref{fig:in_control_weibull_em} reports the results obtained when using the smooth nonparametric baseline excess hazard estimate. The in-control signal ratio obtained is slightly closer to the desired value of 5\%, but the difference in absolute value is still larger compared to the piecewise constant model of above. We suspect that this is a consequence of the choice of the smoothing parameters in the experiment. This example shows a potential issue with the nonparametric baseline excess hazard estimate \cite{perme2009approach} - the smoothing parameters will have a large impact on how the charts will perform compared to the required in-control performance. An ideal situation is to obtain the full continuous estimate of $h_0$ at the last E-step of the EM-algorithm when fitting the Cox-type excess hazard model to avoid another smoothing operation on top of the already smoothed estimates (see e.g. Appendix \ref{appendix:smoothing}). However, this is not possible, and one could argue that this combination might give rise to oversmoothing. Nonetheless, this simulation set-up illustrates the challenges of a nonparametric baseline in the model estimation when using the proposed monitoring scheme. 

\section{Application to colorectal cancer data monitoring}
\label{sec:realdata} 
In this section, we apply the proposed methods to a data set of patients diagnosed with colorectal cancer in Norway between the start of 1970 and until the end of 2020. 
We will focus on patients with specified and known position of the tumor according to the international classification of disease (ICD), who received standard surgical resections for cure of stage I-III colorectal cancer. With reference to Table \ref{table:covariates_sim}, this means that the relevant subset is the patients in surgery group 0, with  SEER stadium equal to localised or regional and with ICD indicator equal to 0, 1 or 2. 
Furthermore, while colorectal cancer is mostly a disease of the elderly, there has been a significant rise in incidence of this disease among the younger population, defined as early-onset colorectal cancer, i.e., $\leq 50$ years of age. Currently, there is an increased focus on this subgroup of colorectal cancer patients (see, e.g. \citeN{collaborative2021characteristics} for an overview). 
Inspired by this fact, we divide the data into patients up to 50 years and patients above 50 years of age. This leaves us with $5321$ patients in the younger age group and $86\,360$ patients in the older age group. 

In this example, we explore monitoring in bands of 10 years, i.e. we choose a baseline period of 10 years and use this period to monitor for the next 10 years. More specifically, we define the first baseline period to be between 1970-1980. The patient cohort diagnosed in this period is used to fit an excess hazard model that will represent $h_{E, 0}(\cdot, X_i)$ in the CUSUM chart. The covariate vector $X_i$ here contains gender, ICD indicator, morphology type and SEER stadium.

In addition, only a piecewise constant baseline excess hazard model is considered here. With these results as the in-control period, we run the CUSUM charts for the next 10 years, i.e. the time period between 1980-1990. Subsequently, this period is then used as the baseline period for the new monitoring period of 1990-2000, and this is done until we monitor the period between 2010 and 2020.

\subsection{Proportional alternative}
\begin{figure}[htbp]
     \centering
     \begin{subfigure}[b]{0.75\textwidth}
         \centering
         \includegraphics[width=\textwidth]{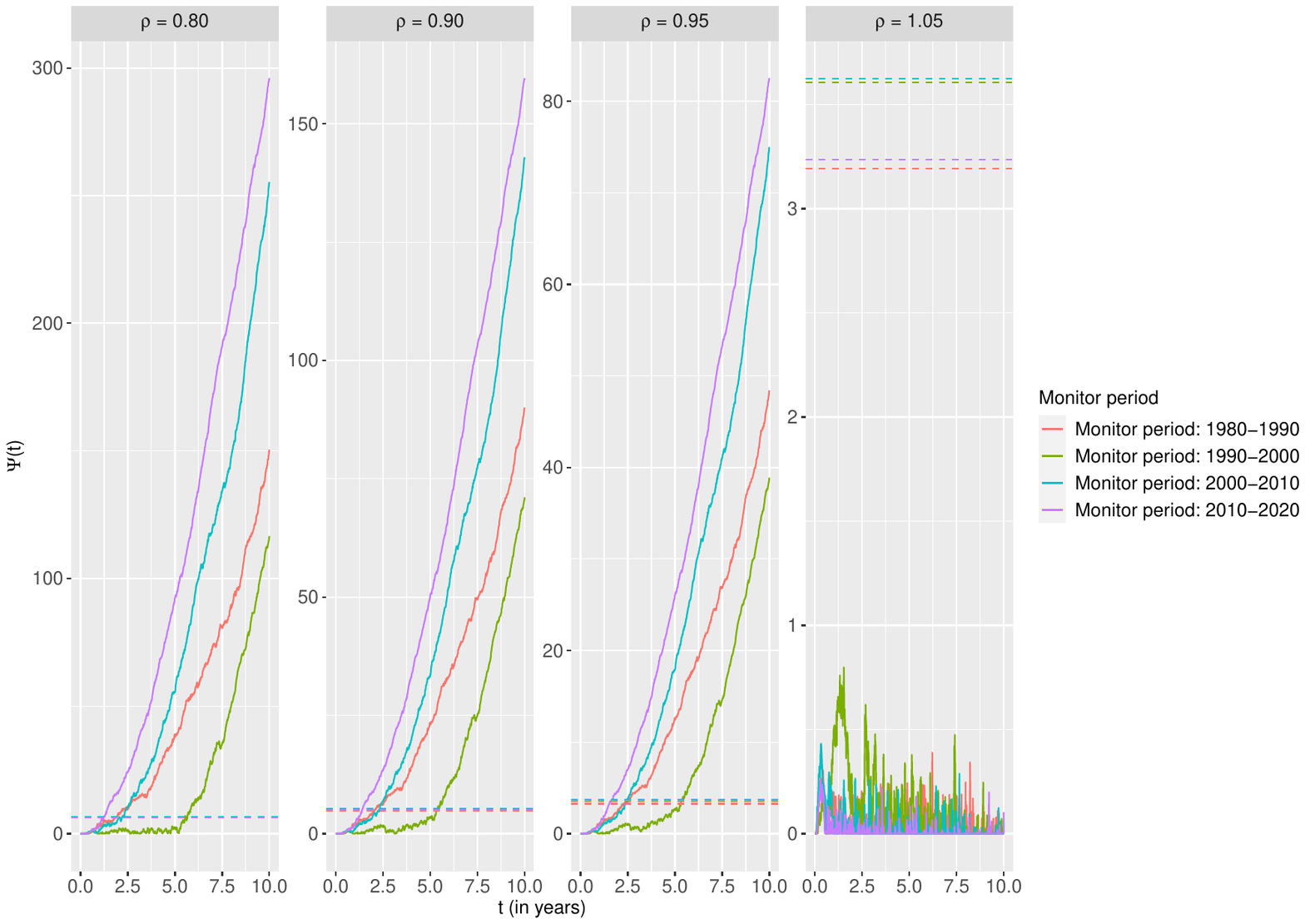}
         \caption{Age $> 50$}
         \label{fig:prop_real_old}
     \end{subfigure}
     \hfill
     \begin{subfigure}[b]{0.75\textwidth}
         \centering
         \includegraphics[width=\textwidth]{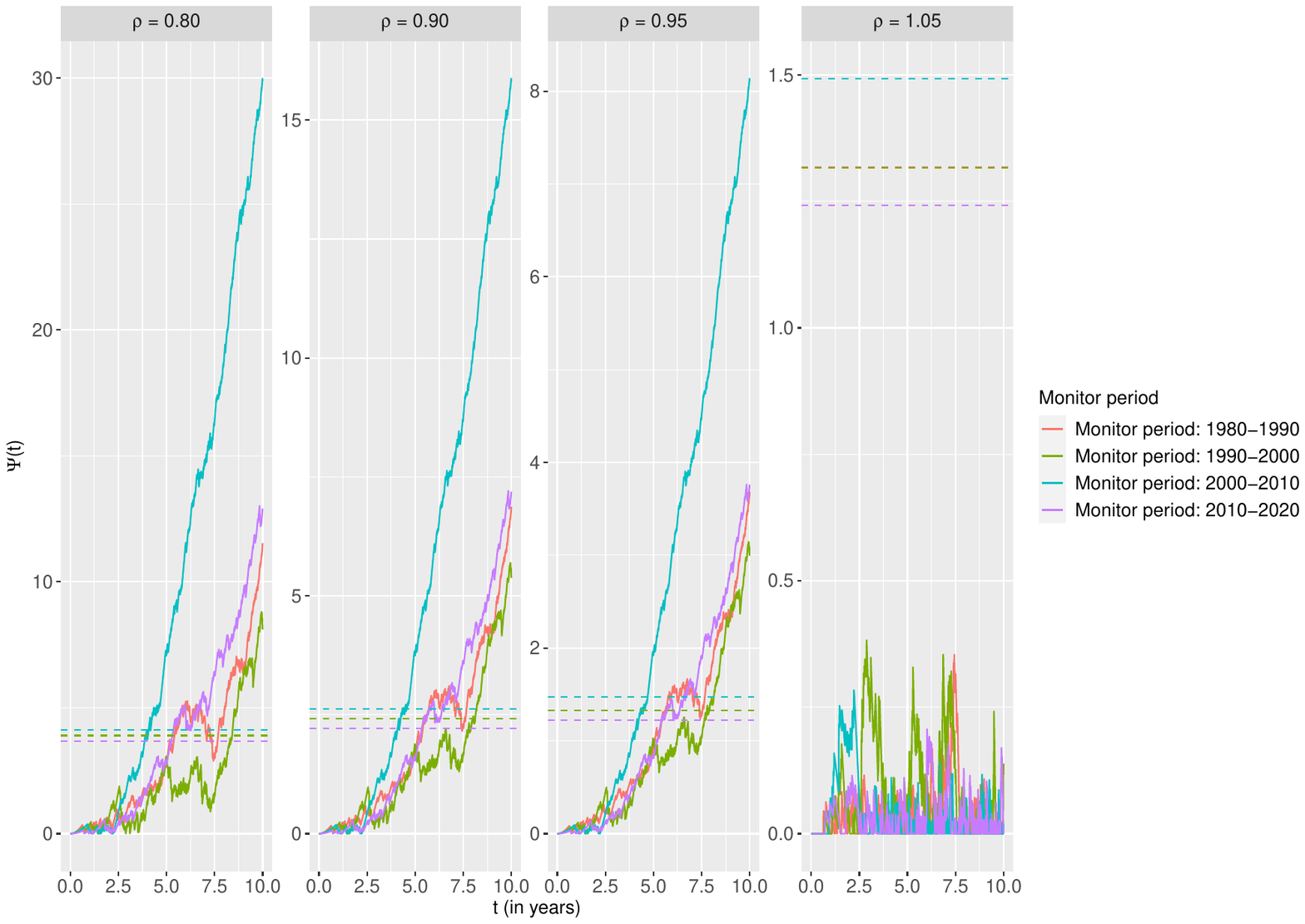}
         \caption{Age $\leq 50$}
         \label{fig:prop_real_young}
     \end{subfigure}
     \caption{CUSUM charts using the proportional alternative for each age group monitoring 10-year survival in four different time periods with the estimated output from the piecewise constant baseline excess hazard model fitted on the preceding 10-year time period as the in-control. Four different values of $\rho$ are explored here.  The dashed lines are the thresholds obtained by simulations using the bootstrap procedure mentioned in Section \ref{subsec:estimation_error}.}
     \label{fig:prop_real}
\end{figure}

First, we consider the proportional alternative. For each combination of age group and monitoring period, the chart is computed using four different values of $\rho$: 0.80, 0.90, 0.95 and 1.05, i.e. three scenarios corresponding to decreased and one to increased excess hazard. Figure \ref{fig:prop_real} shows the calculated charts for different combinations of monitoring time period and $\rho$ in the two age groups. 
 The dashed lines are the thresholds obtained by simulations using the bootstrap procedure mentioned in Section \ref{subsec:estimation_error}. The plots of the charts have different scales for distinct values of $\rho$ as the corresponding CUSUMs are not numerically comparable.


It is clear from both plots that the charts fluctuate and do not signal for any combinations of age group and time period when considering the increased burden of disease scenario of $\rho = 1.05$. In contrast, all charts signal for all the $\rho<1$ scenarios, indicating that improvement versus the previous decade is observed for all the periods considered. For the elders, a striking observation is that in the 1990s, no improvement was seen during the first five years. 
An explanation for this could be that possible advancements in diagnosis and treatment of colon cancer during the 1980s were not reflected in changes in outcomes during the first half of the 1990s. On the other hand, the chart signals fastest in the period 2010-2020. This could indicate early advances in this period, but the faster signal could also possibly be partly explained by more patients arriving during this time interval. For the younger, there seems to be an indication of a better treatment advance in the period 2000-2010 versus the previous decade than in the other periods. The generally later signals for the younger could be explained by much fewer patients in this group.


\subsection{Additive alternative}

\begin{figure}[htbp]
     \centering
     \begin{subfigure}[b]{0.75\textwidth}
         \centering
         \includegraphics[width=\textwidth]{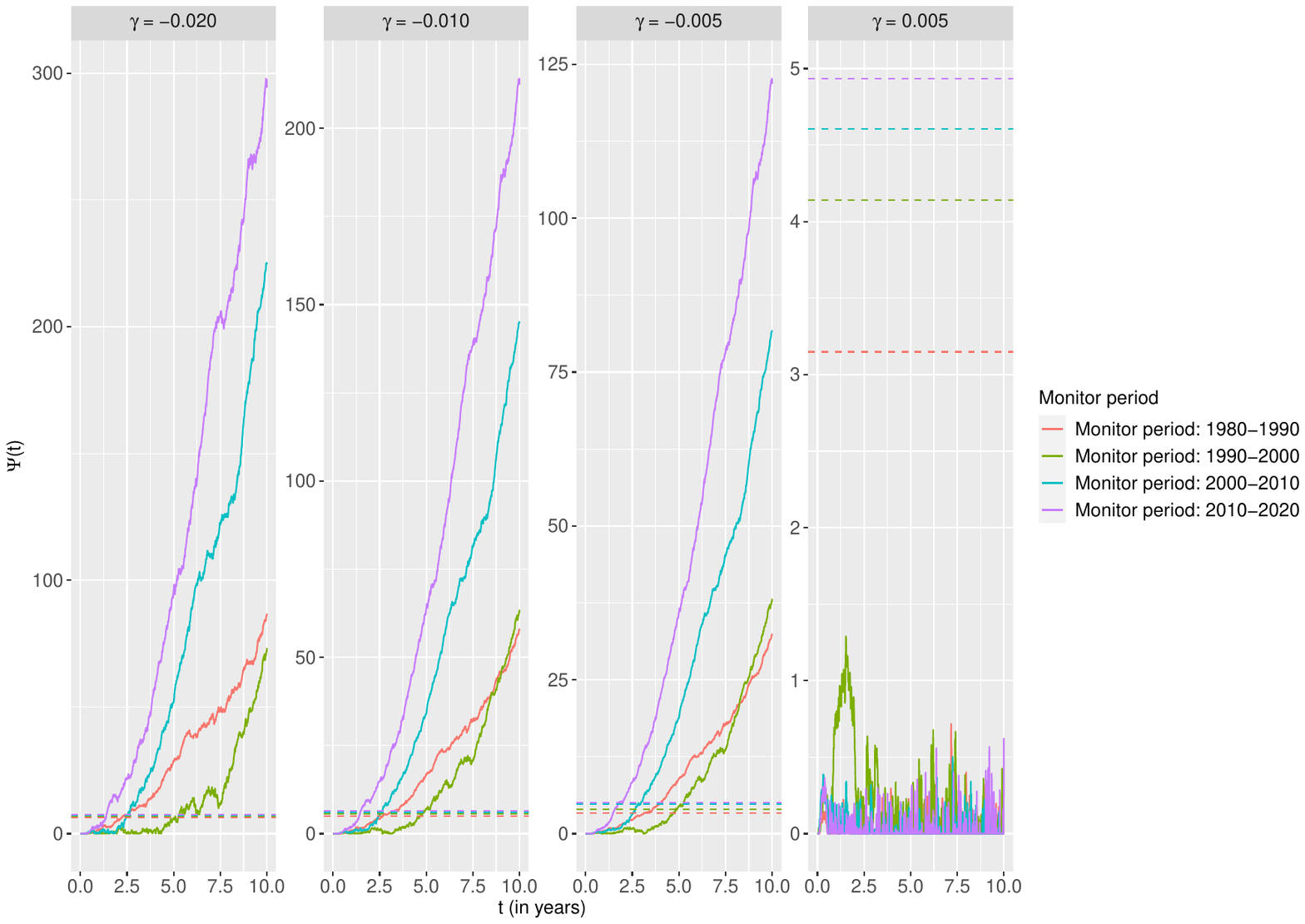}
         \caption{Age $> 50$}
         \label{fig:add_real_old}
     \end{subfigure}
     \hfill
     \begin{subfigure}[b]{0.75\textwidth}
         \centering
         \includegraphics[width=\textwidth]{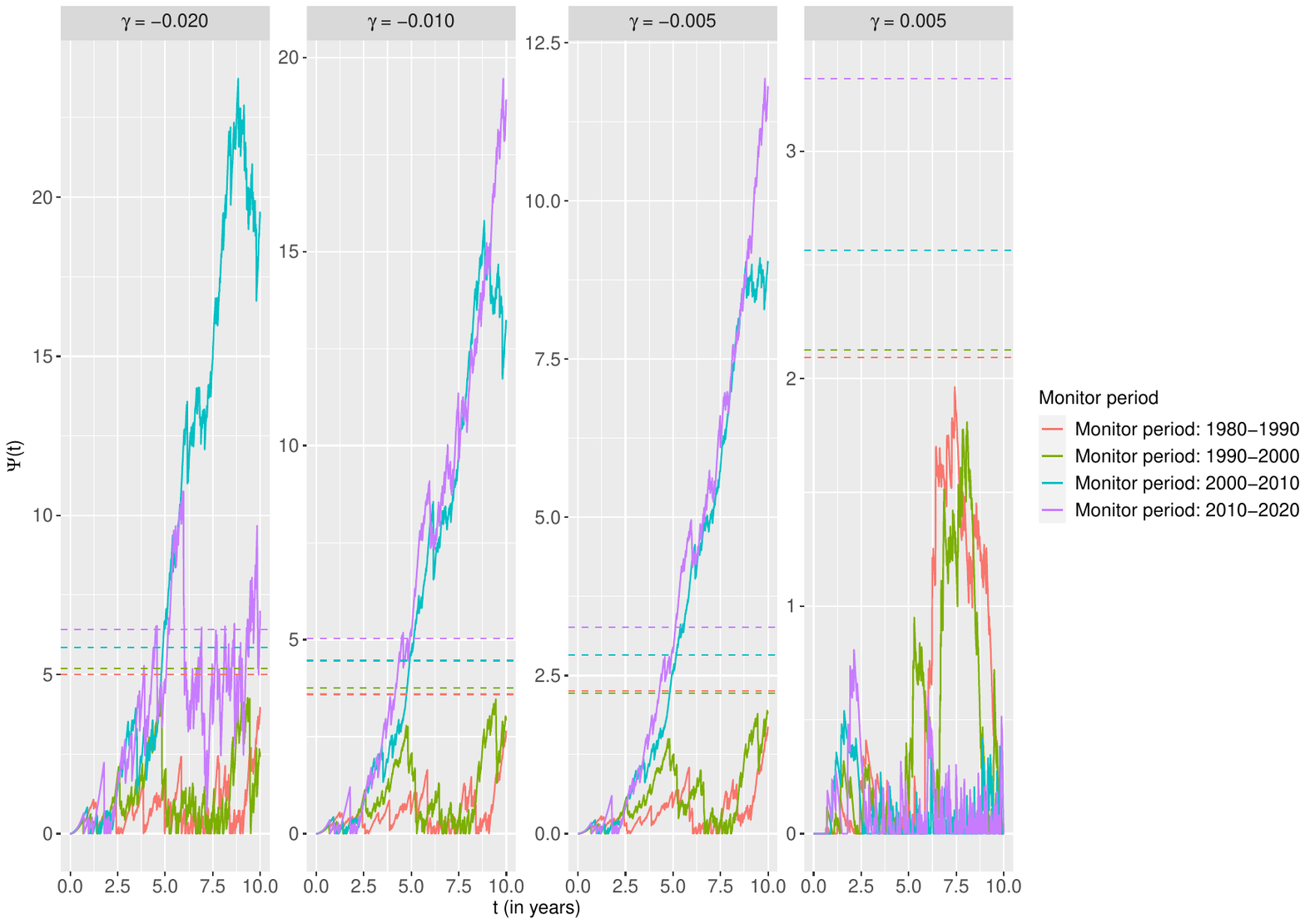}
         \caption{Age $\leq 50$}
         \label{fig:add_real_young}
     \end{subfigure}
     \caption{CUSUM charts using the additive alternative for each age group monitoring 10-year survival in four different time periods with the estimated output from the piecewise constant baseline excess hazard model fitted on the preceding 10-year time period as the in-control. Four different values of $\gamma$ are explored here.  The dashed lines are the thresholds obtained by simulations using the bootstrap procedure mentioned in Section \ref{subsec:estimation_error}.}
     \label{fig:add_real}
\end{figure}

Next, we investigate the setting of an additive alternative. As in the previous setting, we also here use three shift parameters that correspond to decreasing and one corresponding to increasing excess hazard, i.e. the values of $\gamma$ considered are -0.020, -0.010, -0.005 and 0.005. The results are depicted in Figure~\ref{fig:add_real}. We observe that none of the charts  yield a signal for the $\gamma = 0.005$ case that implies a worse outcome over time in this additive setting, although for the younger patients, the charts are actually close to a signal in the periods 1980-1990 and 1990-2000.   


For the remaining values of $\gamma$, all the charts related to the old age group signal with an increasing trend. The behaviors of the charts across all the values of $\gamma$ are very similar to the proportional alternative, with the latest two monitoring periods appearing to have the largest improvement. 

Looking at the charts obtained from the younger age group, no signals are now obtained in the first two monitoring periods for $\gamma < 1$ unlike the conclusions from the proportional alternative. The values of the charts for the periods 2000-2010 and 2010-2020 when $\gamma = -0.010$ and $\gamma = -0.005$ are also much closer. In addition, for the largest shift $\gamma = -0.020$,  the chart for the period 2010-2020 starts decreasing again after signalling at around 5 years. Therefore, the evidence of a larger reduction of $-0.020$ in the excess hazard is much weaker for this period compared to the evidence for smaller changes like $-0.010$ and $-0.005$. This shows the effect of setting a value of the change parameter to a larger change than the observed, and that monitoring for different values of the alternative might give useful insight. 

\subsection{Linear accelerated time alternative}


\begin{figure}[htbp]
     \centering
     \begin{subfigure}[b]{0.75\textwidth}
         \centering
         \includegraphics[width=\textwidth]{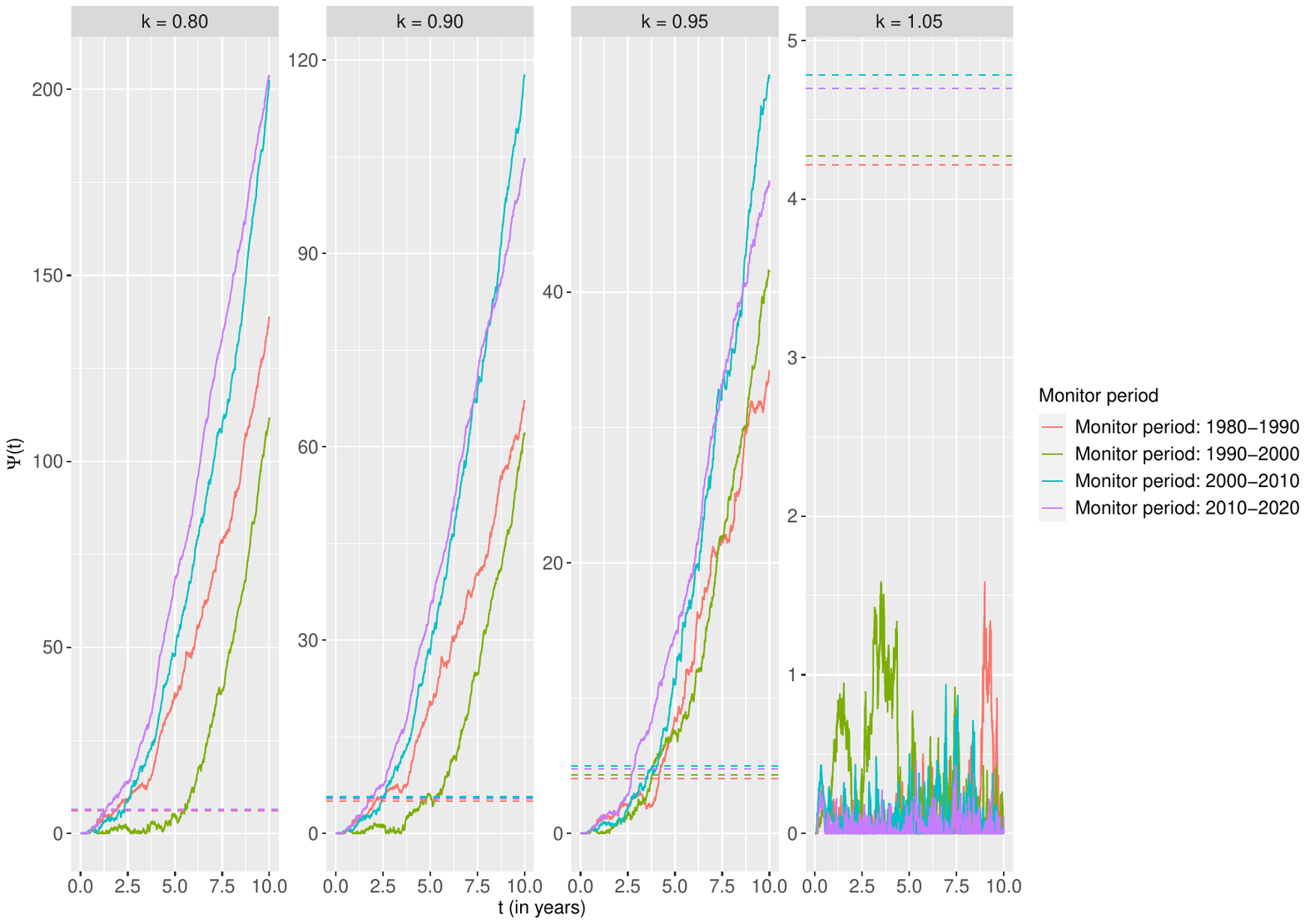}
         \caption{Age $> 50$}
         \label{fig:acc_time_lin_real_old}
     \end{subfigure}
     \hfill
     \begin{subfigure}[b]{0.75\textwidth}
         \centering
         \includegraphics[width=\textwidth]{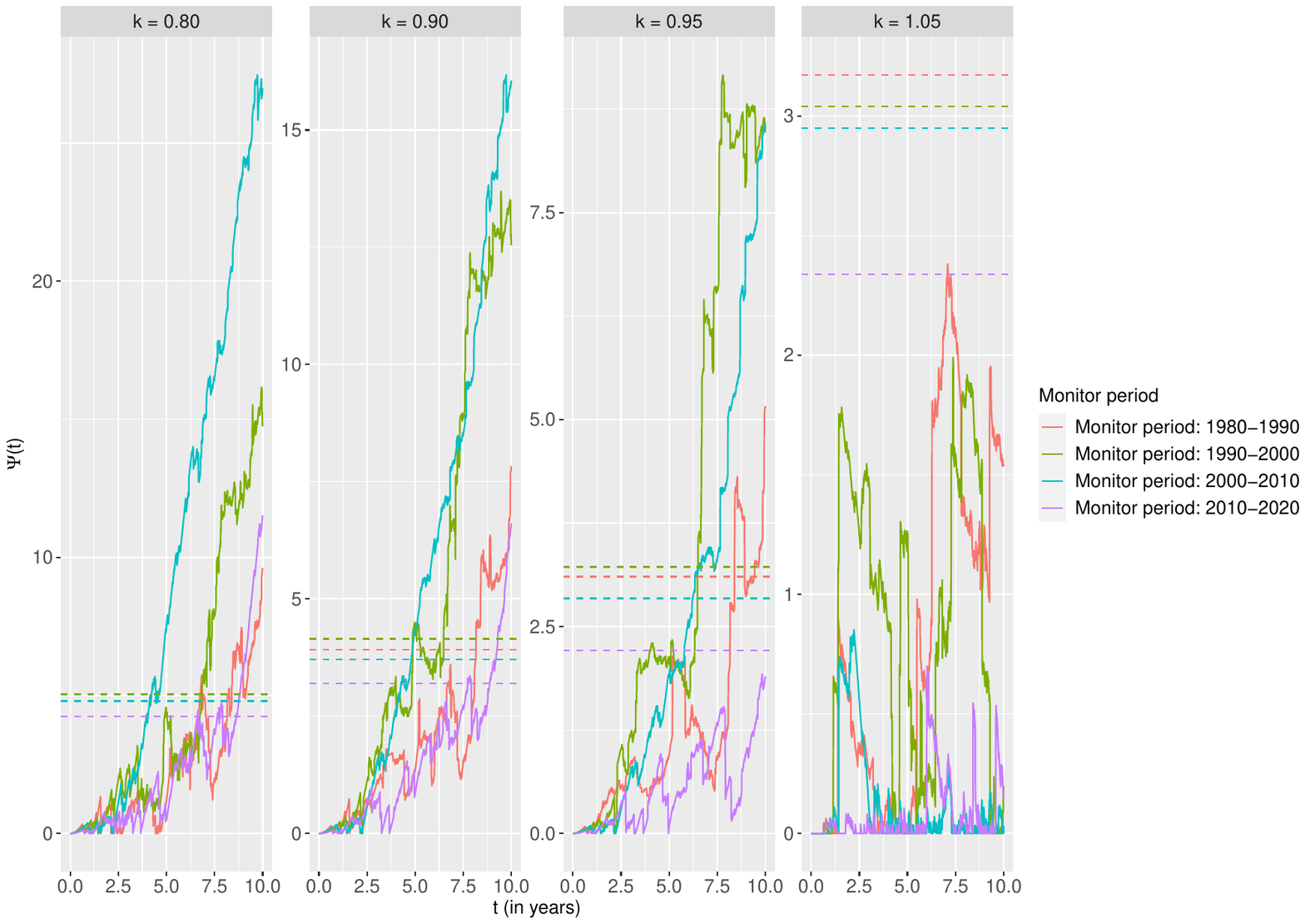}
         \caption{Age $\leq 50$}
         \label{fig:acc_time_lin_real_young}
     \end{subfigure}
     \caption{CUSUM charts using the linear accelerated time alternative for each age group monitoring 10-year survival in four different time periods with the estimated output from the  piecewise constant baseline excess hazard model fitted on the preceding 10-year time period as the in-control. Four different values of $k$ are explored here.  The dashed lines are the thresholds obtained by simulations using the bootstrap procedure mentioned in Section \ref{subsec:estimation_error}.}
     \label{fig:acc_time_lin_real}
\end{figure}


Finally, we try a linear accelerated time alternative with the following four values of the acceleration parameter $k$: 0.80, 0.90, 0.95 and 1.05. 
Therefore, this corresponds again to three improvement alternatives and one worsening situation. 
For the older age group, we see a similar behaviour as in the previous models, although with a later signal in the case with the smallest improvement of $k=0.95$. 


The most interesting observation from the results regarding the younger age group is the charts corresponding to the periods 1990-2000 and 2010-2020. With the additive alternative, the charts monitoring towards improvement for 1990-2000 did not signal at all. However, with the linear accelerated time alternative, all charts for this period signal for all values of $k < 1$, as was also the case for the corresponding charts with the proportional model. Further, the charts for 2010-2020 have a much later response than what was observed for the additive and proportional alternative. Also, when $k = 0.95$, the chart does not even signal. This is another example of how different alternatives can potentially lead to different conclusions. 


\section{Summary}
\label{sec:conc}

We have presented a CUSUM-based method for monitoring changes in excess hazard for relative survival settings where the population hazard is known. This complements the literature on monitoring based on time to event models. The proposed method can be used either for real-time monitoring or for retrospective analyses. The method can be adapted to various data updating schemes.  

We have considered proportional, additive and linear accelerated time change models and studied properties by simulations and in an application to cancer registry data. In particular, we have also considered the impact of estimation error and some forms of model misspecifications. Simulations indicate that model misspecifications might be a somewhat severe issue, while with a decent amount of baseline data the impact of estimation error is not so critical. For small sample sizes, the bootstrap approach for handling estimation error presented in \citeN{GandyKvaloy2013} could in principle be adapted, although with a substantial computational burden in the current setting. Applications to data illustrate that careful considerations need to be made when specifying the type of changes in the excess hazard to monitor against.

Finally, it could be of interest to avoid the requirement of specifying a specific alternative in order to capture more general out-of-control situations. \citeN{phinikettos2014omnibus} have proposed a method that avoids this necessity for the setting of non-risk-adjusted total hazard monitoring. Extending this methodology to the relative survival situation of interest here, but also allowing for the incorporation of covariates via regression models could be an interesting direction in the future.

\bigskip
\bigskip
\bigskip

\noindent
\textbf{Disclaimer}

\noindent
The study has used data from the Cancer Registry of Norway. The interpretation and reporting of these data are the sole responsibility of the authors, and no endorsement by the Cancer Registry of Norway is intended nor should be inferred.

\newpage

\appendix

\section{Summary table of the covariates in the simulation set-up of Section 3} \label{appendix:summary_table}

\begin{table}[htbp]		
		\begin{centering}
			\caption{Overview of the covariates used in the simulations, with the corresponding parameter value used and proportion of observations simulated to be in each category for the categorical covariates. Age is simulated from a normal distribution with mean 75 and standard deviation 10, truncated at 50 and 105. See Tran (2022) for details about the medical interpretations of the covariates.}
               \label{table:covariates_sim}
			\par\end{centering}
		
		\centering{}%
		\begin{adjustbox}{width=0.75\textwidth}
		\begin{tabular}{ccc}
			\hline 
			Variable & $\beta$ & Proportion \tabularnewline
			\hline 
			Gender = Male & $\cdot$ &  50.00\% \tabularnewline
			Gender = Female & $0.005$ & 50.00\% \tabularnewline
			\hline
			ICD indicator = 0 & $\cdot$ &  27.00\% \tabularnewline
			ICD indicator = 1 & $0.500$ &  44.00\% \tabularnewline	
			ICD indicator = 2 & $0.200$ &  28.00\% \tabularnewline
			ICD indicator = 3 & $0.300$ &  1.00\% \tabularnewline
            \hline
           	Morphology type = Adenocarcinoma & $\cdot$ &  90.00\% \tabularnewline
        	Morphology type = Mucinous carcinoma &  $-0.050$ &  10.00\% \tabularnewline
			\hline
           	SEER stadium = Distant & $\cdot$ &  20.00\% \tabularnewline
           	SEER stadium = Localised & $-3.000$ &  20.00\%\tabularnewline
        	SEER stadium = Regional & $-1.750$ &  55.00\% \tabularnewline
        	SEER stadium = Unknown & $-1.000$ &  5.00\% \tabularnewline
        	\hline
			Surgery group = 0 & $\cdot$ &  82.75\% \tabularnewline
			Surgery group = 1 & $1.500$ &  17.15\% \tabularnewline
			Surgery group = 2 & $2.500$ &  0.10\% \tabularnewline
			\hline
		\end{tabular}
		\end{adjustbox}
\end{table}

\section{Smoothing procedure for the semiparametric excess hazard model in the CUSUM method} \label{appendix:smoothing}
An important note regarding the semiparametric excess hazard model implemented in the R package \texttt{relsurv} \cite{perme2018nonparametric} is that the user can specify a smoothing parameter \texttt{bwin}. Originally, the estimation of the model is done using the EM-algorithm by \citeN{dempster1977maximum}. \citeN{perme2009approach} proposed to apply kernel smoothing to the  nonparametric baseline excess hazard estimates at each E-step of the algorithm. The argument \texttt{bwin} therefore controls the bandwidth in the smoothing process, with larger values implying larger bandwidth, and therefore smoother estimates. In the end, only the estimates of $h_0$ at the observed times to event are returned. However, in order to apply the proposed CUSUM procedures, $h_0(t)$ is required at all possible values of $t \in [0, 10]$ for continuous monitoring if the main interest is the 10-year excess survival. To get a continuous estimate of $h_0(t)$, we must again apply some sort of manual smoothing on top of the returned estimates at the times to event observed in the data. For this purpose, smoothing splines (see e.g. \citeN{james2013introduction}) are used with the baseline excess hazard estimates as the response variable and the times to event as the sole predictor. Therefore, there are two extra sources that can impact the upcoming results: The smoothing parameter \texttt{bwin} at each E-step of the EM-algorithm when fitting the model and the effective degrees of freedom from the smoothing splines procedure. To get somewhat sensible results, we have decided to set \texttt{bwin} and the effective degrees of freedom to 25. It is therefore important to note that a different combination of these two parameters will lead to a different result than what we have obtained in Section \ref{subsec:estimation_error}. 

\bibliography{EHCbib}

@STRING{Annals = {The Annals of Statistics}}

@STRING{BIOMETRICS = {Biometrics}}

@STRING{BIOMETRIKA = {Biometrika}}

@article{Gandy2010ram,
  title={Risk-Adjusted Monitoring of Time to Event},
  author={Gandy, A. and Kval{\o}y, J.T. and Bottle, A. and Zhou, F.},
  journal={Biometrika},
  volume={97},
  number={2},
  year={2010},
   pages={375--388}
}

@article{Moustakides1986,
  title={{Optimal Stopping Times for Detecting Changes in Distributions}},
  author={Moustakides, G.V.},
  journal={Annals of Statistics},
  volume={14},
  number={4},
  year={1986},
   pages={1379--1387}
}

@Book{abg2008,
  author = 	 {Aalen, O. O. and Borgan, {\O} and Gjessing, H. K.},
  title = 	 {Survival and Event History Analysis: A Process Point of View.},
  publisher = {New York: Springer},
  year = 	 2008
}

@MastersThesis{Tran2022,
  author = 	 {Tran, Jimmy .H.},
  title = 	 {{Relative Survival Methods. Theory, Applications and Extensions to Monitoring}},
  school = 	 {Department of Mathematics and Physics, University of Stavanger}, 
  year = 	 2022
}

@article{dickman2004regression,
  title={Regression models for relative survival},
  author={Dickman, Paul W and Sloggett, Andy and Hills, Michael and Hakulinen, Timo},
  journal={Statistics in Medicine},
  volume={23},
  number={1},
  pages={51--64},
  year={2004},
  publisher={Wiley Online Library}
}

@article{perme2009approach,
  title={An approach to estimation in relative survival regression},
  author={Perme, Maja Pohar and Henderson, Robin and Stare, Janez},
  journal={Biostatistics},
  volume={10},
  number={1},
  pages={136--146},
  year={2009},
  publisher={Oxford University Press}
}

@article{perme2018nonparametric,
  title={Nonparametric {R}elative {S}urvival {A}nalysis with the {R} {P}ackage relsurv},
  author={Perme, Maja Pohar and Pavlic, Klemen},
  journal={Journal of Statistical Software},
  volume={87},
  pages={1--27},
  year={2018}
}

@article{dempster1977maximum,
  title={{Maximum likelihood from incomplete data via the EM algorithm}},
  author={Dempster, Arthur P and Laird, Nan M and Rubin, Donald B},
  journal={Journal of the Royal Statistical Society: Series B (Methodological)},
  volume={39},
  number={1},
  pages={1--22},
  year={1977},
  publisher={Wiley Online Library}
}

@book{james2013introduction,
  title={An Introduction to Statistical Learning},
  author={James, Gareth and Witten, Daniela and Hastie, Trevor and Tibshirani, Robert},
  volume={112},
  year={2013},
  publisher={Springer}
}

@article{Liu2023SCRCUSUM,
title = {{SCR-CUSUM: An illness-death semi-Markov model-based risk-adjusted CUSUM for semi-competing risk data monitoring}},
journal = {Computers \& Industrial Engineering},
volume = {184},
pages = {109530},
year = {2023},
issn = {0360-8352},
doi = {https://doi.org/10.1016/j.cie.2023.109530},
author = {Ruoyu Liu and Xin Lai and Jiayin Wang and Xiaoyan Zhu and Yuqian Liu},
}

@article{Begun2019CUSUMfrailty,
title = {Risk-adjusted {CUSUM} control charts for
shared frailty survival models with
application to hip replacement outcomes: a
study using the {NJR} dataset},
journal = {BMC Medical Research Methodology},
volume = {19},
year = {2019},
doi = {https://doi.org/10.1186/s12874-019-0853-2},
author = {Alexander Begun and Elena Kulinskaya and Alexander J MacGregor},
}

@article{Sego2009Riskadjusted,
author = {Sego, Landon H. and Reynolds Jr, Marion R. and Woodall, William H.},
title = {Risk-adjusted monitoring of survival times},
journal = {Statistics in Medicine},
volume = {28},
number = {9},
pages = {1386-1401},
doi = {https://doi.org/10.1002/sim.3546},
year = {2009}
}

@article{Biswas2008Riskadjusted,
author = {Biswas, Pinaki and Kalbfleisch, John D.},
title = {{A risk-adjusted CUSUM in continuous time based on the Cox model}},
journal = {Statistics in Medicine},
volume = {27},
number = {17},
pages = {3382-3406},
doi = {https://doi.org/10.1002/sim.3216},
year = {2008}
}

@article{Zhang2016Estimationerror,
author = {Zhang, Min and Xu, Yahui and He, Zhen and Hou, Xuejun},
title = {{The Effect of Estimation Error on Risk-Adjusted Survival Time CUSUM Chart Performance}},
journal = {Quality and Reliability Engineering International},
volume = {32},
number = {4},
pages = {1445-1452},
doi = {https://doi.org/10.1002/qre.1849},
year = {2016}
}

@article{Oliveira2016Longterm,
author = {Oliveira, Jocelânio W. and Valença, Dione M. and Medeiros, Pledson G. and Marçula, Magaly},
title = {Risk-adjusted monitoring of time to event in the presence of long-term survivors},
journal = {Biometrical Journal},
volume = {58},
number = {6},
pages = {1485-1505},
doi = {https://doi.org/10.1002/bimj.201500094},
year = {2016}
}

@article{Chen2023CUSUMperiopmortality,
    author = {Chen, Vivi W. and Chidi, Alexis P. and Dong, Yongquan and Richardson, Peter A. and Axelrod, David A. and Petersen, Laura A. and Massarweh, Nader N.},
    title = {{Risk-Adjusted Cumulative Sum for Early Detection of Hospitals With Excess Perioperative Mortality}},
    journal = {JAMA Surgery},
    volume = {158},
    number = {11},
    pages = {1176-1183},
    year = {2023},
    month = {11},
    issn = {2168-6254},
    doi = {10.1001/jamasurg.2023.3673}
}

@article{Lai2021CUSUMsurgerysurvival,
title = {A risk-adjusted approach to monitoring surgery for survival outcomes based on a weighted score test},
journal = {Computers \& Industrial Engineering},
volume = {160},
pages = {107568},
year = {2021},
issn = {0360-8352},
doi = {https://doi.org/10.1016/j.cie.2021.107568},
author = {Xin Lai and Xiao Li and Liu Liu and Fugee Tsung and Paul B.S. Lai and Jiayin Wang and Xuanping Zhang and Xiaoyan Zhu and Jiaqi Liu} 
}

@article{Kuang2023CUSUMqueues,
author = {Yanqing Kuang and Devashish Das and Mustafa Sir and Kalyan Pasupathy},
title = {{Likelihood ratio-based CUSUM charts for real-time monitoring the quality of service in a network of queues}},
journal = {IISE Transactions on Healthcare Systems Engineering},
volume = {13},
number = {4},
pages = {344--354},
year = {2023},
publisher = {Taylor and Francis},
doi = {10.1080/24725579.2023.2181470}
}

@article{Massarweh2021,
author = {Nader N. Massarweh and Vivi W. Chen and Tracey Rosen and Yongquan Dong and Peter A. Richardson and David A. Axelrod and  Alex H.S. Harris and Mark A. Wilson and Laura A. Petersen},
title = {{Comparative Effectiveness of Risk-adjusted Cumulative Sum and Periodic Evaluation for Monitoring Hospital Perioperative Mortality}},
journal = {Medical Care},
volume = {59},
number = {7},
pages = {639--645},
year = {2021},
doi = {10.1097/MLR.0000000000001559}
}

@article{Keshavarz2021CUSUMfrailty,
author = {Maryam Keshavarz and Shervin Asadzadeh and Seyed Taghi Akhavan Niaki},
title = {{Risk-adjusted frailty-based CUSUM control chart for phase I monitoring of patients' lifetime}},
journal = {Journal of Statistical Computation and Simulation},
volume = {91},
number = {2},
pages = {334--352},
year = {2021},
publisher = {Taylor and Francis},
doi = {10.1080/00949655.2020.1814775}
}

@article{Steiner2010EWMAsurv,
author = {Steiner, Stefan H. and Jones, Mark},
title = {Risk-adjusted survival time monitoring with an updating exponentially weighted moving average ({EWMA}) control chart},
journal = {Statistics in Medicine},
volume = {29},
number = {4},
pages = {444-454},
doi = {https://doi.org/10.1002/sim.3788},
year = {2010}
}

@article{collaborative2021characteristics,
  title={{Characteristics of Early-Onset vs Late-Onset Colorectal Cancer: A Review}},
  author={{REACCT Collaborative} and Zaborowski, Alexandra M and Abdile, Ahmed and Adamina, Michel and Aigner, Felix and others},
  journal={JAMA surgery},
  volume={156},
  number={9},
  pages={865--874},
  year={2021},
  publisher={American Medical Association}
}

@article{perme2012estimation,
  title={{On Estimation in Relative Survival}},
  author={Perme, Maja Pohar and Stare, Janez and Est{\`e}ve, Jacques},
  journal={Biometrics},
  volume={68},
  number={1},
  pages={113--120},
  year={2012},
  publisher={Oxford University Press}
}

@article{page1954continuous,
  title={{Continuous Inspection Schemes}},
  author={Page, Ewan S},
  journal={Biometrika},
  volume={41},
  number={1/2},
  pages={100--115},
  year={1954},
  publisher={JSTOR}
}

@article{phinikettos2014omnibus,
  title={{An omnibus CUSUM chart for monitoring time to event data}},
  author={Phinikettos, Ioannis and Gandy, Axel},
  journal={Lifetime Data Analysis},
  volume={20},
  pages={481--494},
  year={2014},
  publisher={Springer}
}

@article{dickman2015Stata,
  title={{Estimating and Modeling Relative Survival}},
  author={Dickman, Paul W. and Coviello, Enzo},
  journal={The Stata Journal},
  volume={15},
  pages={186--215},
  year={2015},
}

@article{GandyKvaloy2013,
author = {Gandy, Axel and Kvaløy, Jan Terje},
title = {{Guaranteed Conditional Performance of Control Charts via Bootstrap Methods}},
journal = {Scandinavian Journal of Statistics},
volume = {40},
number = {4},
pages = {647-668},
doi = {https://doi.org/10.1002/sjos.12006},
year = {2013}
}

\bibliographystyle{dcu}

\end{document}